\documentclass[preprint2]{aastex}
\usepackage{graphicx}
\usepackage{natbib}
\usepackage{graphics}

\begin{document}

\title{Submillimeter and molecular views of three Galactic ring-like \ion{H}{2} regions}

\author{K. Arvidsson}
\affil{Astronomy Department, Adler Planetarium, 1300 South Lake Shore Drive, Chicago, IL 60605, USA; karvidsson@adlerplanetarium.org}

\author{C. R. Kerton}
\affil{Department of Physics \& Astronomy, Iowa State University, Ames, IA 50011, USA; kerton@iastate.edu}

\begin{abstract}
We use SCUBA 850 $\mu$m and CO observations to analyze the surroundings of three Galactic ring-like \ion{H}{2} regions, KR~7, KR~81 and KR~120 (Sh 2-124, Sh 2-165 and Sh 2-187), with the aim of finding sites of triggered star formation. We find one prominent submillimeter (sub-mm) source for each region, located at the interface between the \ion{H}{2} region and its neutral surroundings. Using Two Micron All Sky Survey photometry, we find that the prominent sub-mm source for KR~120 probably contains an embedded cluster of young stellar objects (YSOs), making it a likely site for triggered star formation. The KR~7 sub-mm source could possibly contain embedded YSOs, while the KR~81 sub-mm source likely does not. The mass column densities for these dominant sub-mm sources fall in the $\sim 0.1 - 0.6$ g~cm$^{-2}$ range. The mass of the cold, dense material (clumps) seen as the three dominant sub-mm sources fall around $\sim 100$ $M_{\sun}$. We use the SCUBA Legacy catalog to characterize the populations of sub-mm sources around the \ion{H}{2} regions, and compare them to the sources found around a previously studied similar ring-like \ion{H}{2} region (KR~140) and near a massive star-forming region (W3). Finally, we estimate the IR luminosities of the prominent newly detected sub-mm sources and find that they are correlated with the clump mass, consistent with a previously known luminosity-mass relationship which this study shows to be valid over four orders of magnitude in mass.
\end{abstract}


\section{Introduction}
Young stellar objects (YSOs) are often observed at the peripheries of \ion{H}{2} regions, suggesting that star formation can be triggered at such locations \citep{elmegreen-triggeredstarformation-1998ASPC..148..150E,2005A&A...433..565D}. This study is aimed at investigating such sites of possible triggered star formation around a sample of ring-like \ion{H}{2} regions. These \ion{H}{2} regions have a distinctive ring morphology in \emph{Midcourse Space Experiment} Galactic Plane Survey \citep[\emph{MSX};][]{MSX-2001AJ....121.2819P} A-band ($\lambda_0 = 8.3$ $\mu$m, FWHM $\Delta \lambda = 7-11$ $\mu$m) images filled with radio continuum emission. \emph{MSX} A-band emission is dominated by intense line emission from polycyclic aromatic hydrocarbons (PAHs) that arises in the interface between ionized gas in the \ion{H}{2} region and the surrounding molecular material. Inside the \ion{H}{2} region the flux of ultraviolet (UV) photons is sufficiently high to destroy PAHs. Outside of the PAH-free region, in the photodissociation regions, the less energetic UV photon flux excites PAHs but does not destroy them. Moving farther into the cooler molecular gas the UV flux decreases rapidly and PAHs are likely incorporated into larger dust grains. The result is a striking ring-like or ``halo'' morphology  the PAH emission traces an interface region where new stars are possibly being formed.

\begin{figure}
\epsscale{1.3}
\plotone{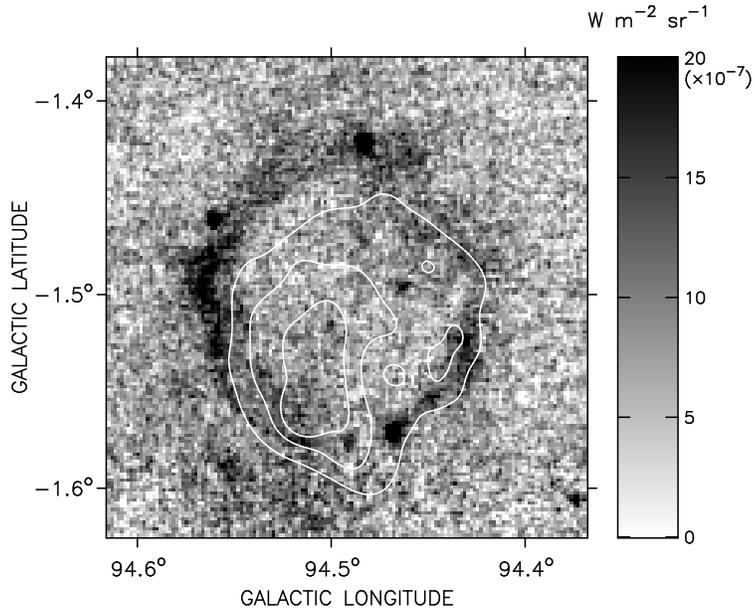}
\caption{\emph{MSX} 8.3 $\mu$m image of KR~7 (Sh 2-124). The white contours correspond to CGPS 1420 MHz continuum brightness temperature levels of 11, 13, and 15 K.}
\label{figure-kr7-msxwithradiocontours}
\end{figure}

\begin{figure}
\plotone{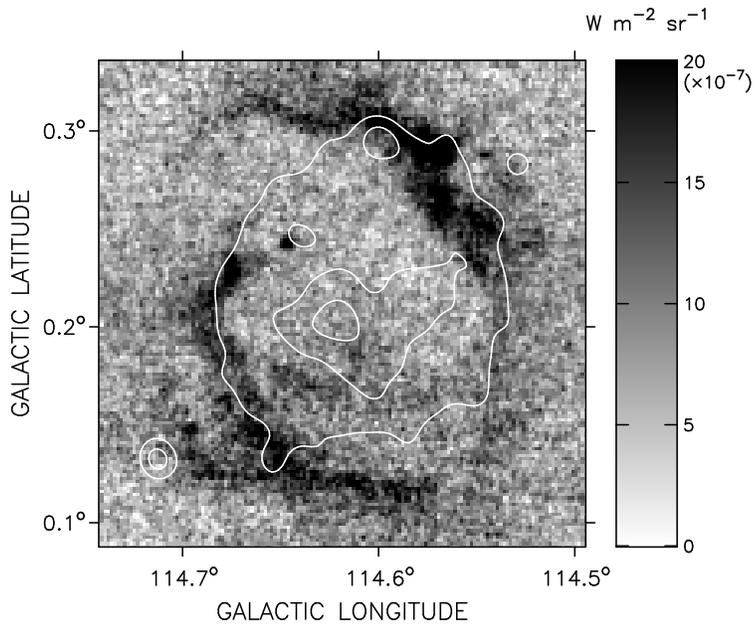}
\caption{\emph{MSX} 8.3 $\mu$m image of KR~81 (Sh 2-165). The white contours correspond to CGPS 1420 MHz continuum brightness temperature levels of 7, 8, and 9 K.}
\label{figure-kr81-msxwithradiocontours}
\end{figure}

\begin{figure}
\plotone{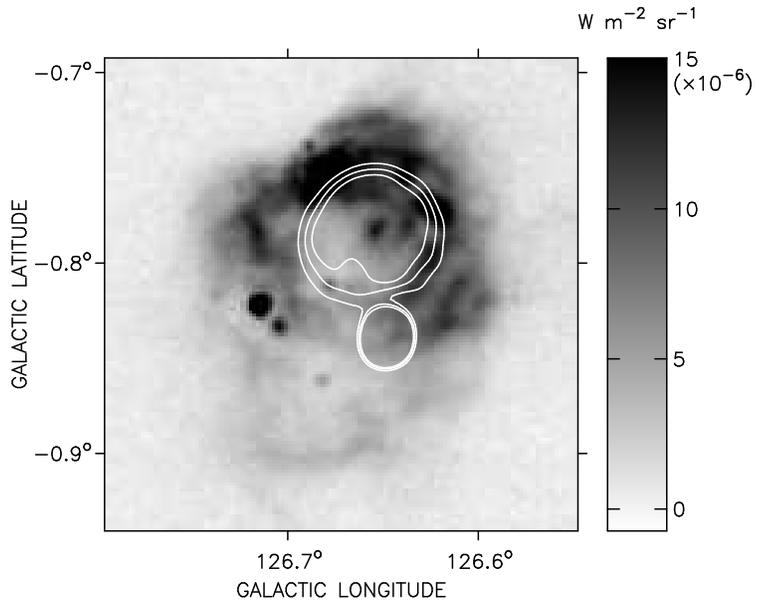}
\caption{\emph{MSX} 8.3 $\mu$m image of KR~120 (Sh 2-187). The white contours correspond to CGPS 1420 MHz continuum brightness temperature levels of 10, 12, and 14 K. The radio source seen at the edge of the main \ion{H}{2} region contours is an extragalactic background source.}
\label{figure-kr120-msxwithradiocontours}
\end{figure}

This study means to address the following issues. What are the physical conditions such as mass and column density at the sites of possible star formation? How many such sites are there? What is the YSO content? This study will look at submillimeter (sub-mm) and molecular line observations toward a selected sample of three ring-like \ion{H}{2} regions, KR~7 (Sh 2-124), KR~81 (Sh 2-165), and KR~120 (Sh 2-187) \citep{sharpless1959ApJS....4..257S,kr80}. These single star powered regions were chosen to minimize the effects of feedback from multiple massive stars. They were selected based on their appearance in the Canadian Galactic Plane Survey \citep[CGPS;][]{tayl03} and \emph{MSX}, displaying the expected roughly circular radio morphology, a ring-like infrared morphology, and a small angular size. Figures~\ref{figure-kr7-msxwithradiocontours}, \ref{figure-kr81-msxwithradiocontours} and \ref{figure-kr120-msxwithradiocontours} show \emph{MSX} A-band images ($\sim 18 \arcsec$ resolution) of the ring-like \ion{H}{2} regions KR~7, KR~81, and KR~120, respectively, overlaid with 1420 MHz radio continuum contours ($\sim 1 \arcmin$ resolution). For KR~120, the radio source seen at the edge of the main \ion{H}{2} region contours is an extragalactic background source. Positions and various other designations of these \ion{H}{2} regions are listed in Table~\ref{table-KRsubmmobservations}.

In addition, these new data will be combined with data on KR~140, another small ring-like \ion{H}{2} region that has been the subject of several studies \citep{kr140-1st-1999AJ....117.2485K,kr140-2nd-2000ApJ...539..283B,kr140-submm-2001ApJ...552..601K,kr140-2008MNRAS.385..995K}. This larger sample will be contrasted with sub-mm observations of a very different environment, the interface between the massive star-forming complex W3 and its surrounding molecular cloud.

\section{Observations}
\subsection{Sub-mm observations}
The sub-mm observations were obtained using the Submillimeter Common-User Bolometer Array (SCUBA) on the 15 m James Clerk Maxwell Telescope (JCMT) at Mauna Kea, Hawaii. SCUBA was used to acquire 450 $\mu$m and 850 $\mu$m images (at $\sim 8 \arcsec$ and $\sim 15 \arcsec$ resolution respectively) during 11 nights in 2003, August 12, 15, and 17, September 17 and 24-26, and October 4 and 9-11.

The reduction was performed following the SCUBA map reduction cookbook\footnote{http://docs.jach.hawaii.edu/star/sc11.htx/sc11.html}. Standard commands within the SCUBA reduction package \textsc{surf} were used to apply flat fields, correct for the signal attenuation in the atmosphere, remove spikes, remove base lines, identify noisy bolometers and remove sky noise. For the observations at 450 $\mu$m, the weather was too poor and no sources could be seen in the images. These 450 $\mu$m observations have been excluded from further study. The individual reduced 850 $\mu$m images were calibrated into Jy~beam$^{-1}$ using the primary calibration source Uranus. To account for the varying background, the reduced and calibrated images were smoothed to a resolution of $120 \arcsec$. The smoothed images were then subtracted from the reduced and calibrated images, thus removing large-scale structure ($>120 \arcsec$). The background-subtracted images of the same target were then co-added to reduce the noise level. The resulting final background-subtracted images were then used in the subsequent analysis. They have noise levels of $0.07 - 0.04$ Jy~beam$^{-1}$.

Figures~\ref{figure-kr7-finalsubmmimage}, \ref{figure-kr81-finalsubmmimage} and \ref{figure-kr120-finalsubmmimage} show the final 850 $\mu$m images of KR~7, KR~81, and KR~120, respectively. By visual examination, they each have one prominent 850 $\mu$m source with a signal-to-noise ratio (S/N) greater than 6 that is located on the ``rim'' seen in the \emph{MSX} images. Faint (S/N $\lesssim 4$) extended sub-mm emission can sometimes be seen that usually lines up with \emph{MSX} emission.

The SCUBA Legacy Catalogues paper \citep{scubalegacy-2008ApJS..175..277D} provides an archive of SCUBA observations, including ours, reduced using a different method \citep[the ``matrix inversion'' method described in][]{johnstone2000ApJS..131..505J}. From these images, catalogs of sources have been extracted using an automated object identification program based on the \textsc{clumpfind} algorithm. By cross-checking the results of the two different reduction techniques we are able to increase the number of detected sub-mm sources around the \ion{H}{2} regions.

\subsection{Molecular observations}
Observations of CO ($J = 1 \rightarrow 0$) line emission from the isotopes $^{12}$CO (115.3 GHz), $^{13}$CO (110.2 GHz) and C$^{18}$O (109.8 GHz) were obtained using the SEcond QUabbin Optical Imaging Array (SEQUOIA) at the FCRAO 14 m telescope in Massachusetts. The reduction, calibration and processing of the raw data into science-ready data cubes was done by C. Brunt in 2004. The molecular line cubes have a spectral resolution of $0.25$ km~s$^{-1}$ and an angular resolution of $46 \arcsec$. The flux scale uncertainty is taken to be $\sim 15 \%$ from \citet{jackson2006ApJS..163..145J}, who used the same instrument to conduct the Galactic Ring Survey. 

\section{Analysis}\label{sec-submmanalysis}

\subsection{KR~7}\label{subsec-kr7analysis}

KR~7 (Sh 2-124) is a $12 \arcmin$ diameter \ion{H}{2} region with a 1420 MHz flux density $F_{1420} = 2690 \pm 81$ mJy \citep{kerton-krcatalogrevisted-2006MNRAS.373.1203K}. Using a spherical, constant density model, a lower limit on the total emission rate of ionizing photons ($N_L$) can be derived from the observed radio flux density
\begin{equation}{\label{eq:N_L}}
N_L \geq 7.5 \times 10^{43} F_{\nu} d^2 \nu^{0.1} T_e^{-0.45} ~\textrm{s}^{-1}\\
\end{equation}
where $\nu$ is the frequency in GHz, $F_{\nu}$ is the flux density measured at frequency $\nu$ in mJy, $d$ is the distance to the source in kpc, and $T_e$ is the electron temperature in units of $10^4$ K \citep{rud96}. The flux of ionizing photons, $\log{N_L} = 48.2$, corresponds to a O9~V star \citep{crowther2005IAUS..227..389C}. \citet{crampton1978A&A....66....1C} classify the possible exciting star of KR~7 (LS III $+50 \degr 24$) as a B0~V star. However, they note that it is not located in the brightest part of the nebula and \citet{lahulla-kr7star-1985A&AS...61..537L} concludes that LS III $+50 \degr 24$ is too cool, leaving the identity of the exciting star for KR~7 an open question.

\begin{figure}
\plotone{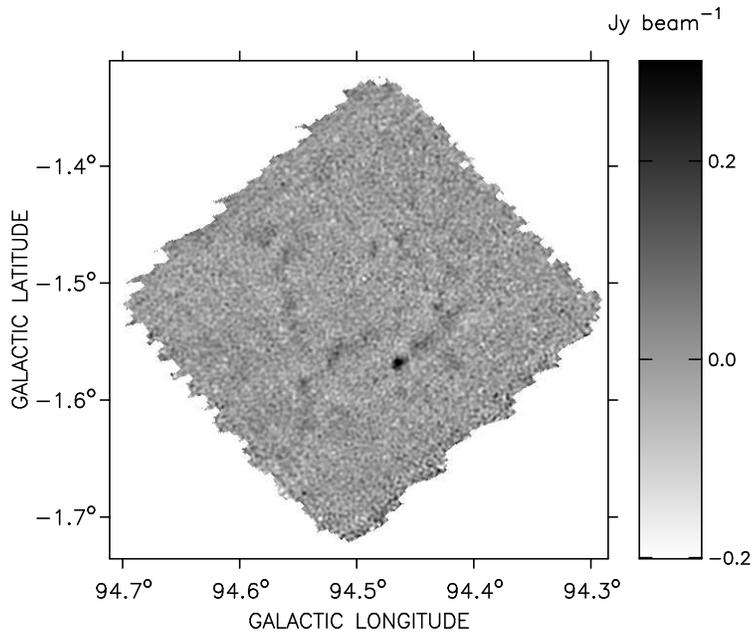}
\caption{Final 850 $\mu$m image of KR~7.}
\label{figure-kr7-finalsubmmimage}
\end{figure}

\begin{figure}
\epsscale{1.4}
\plotone{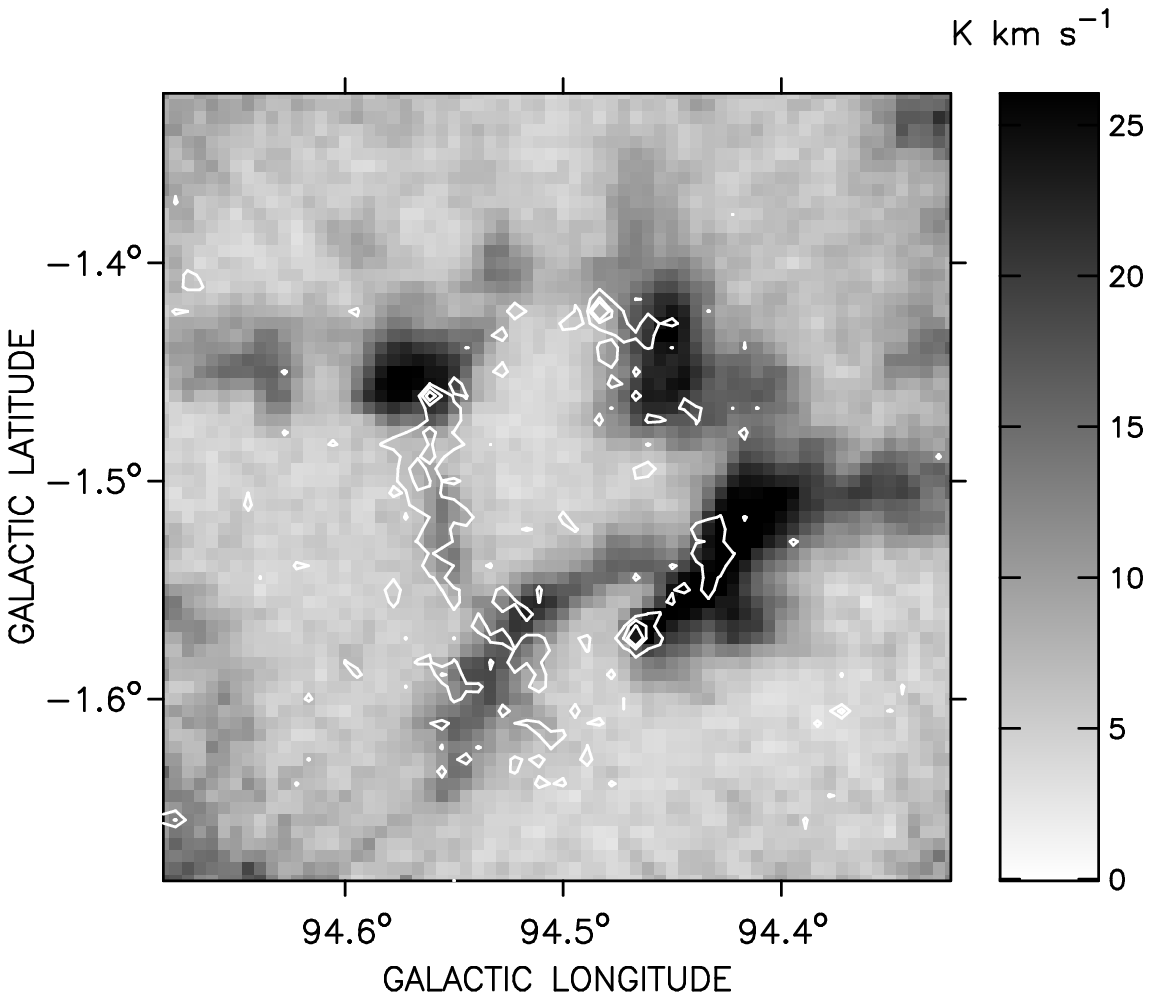}
\plotone{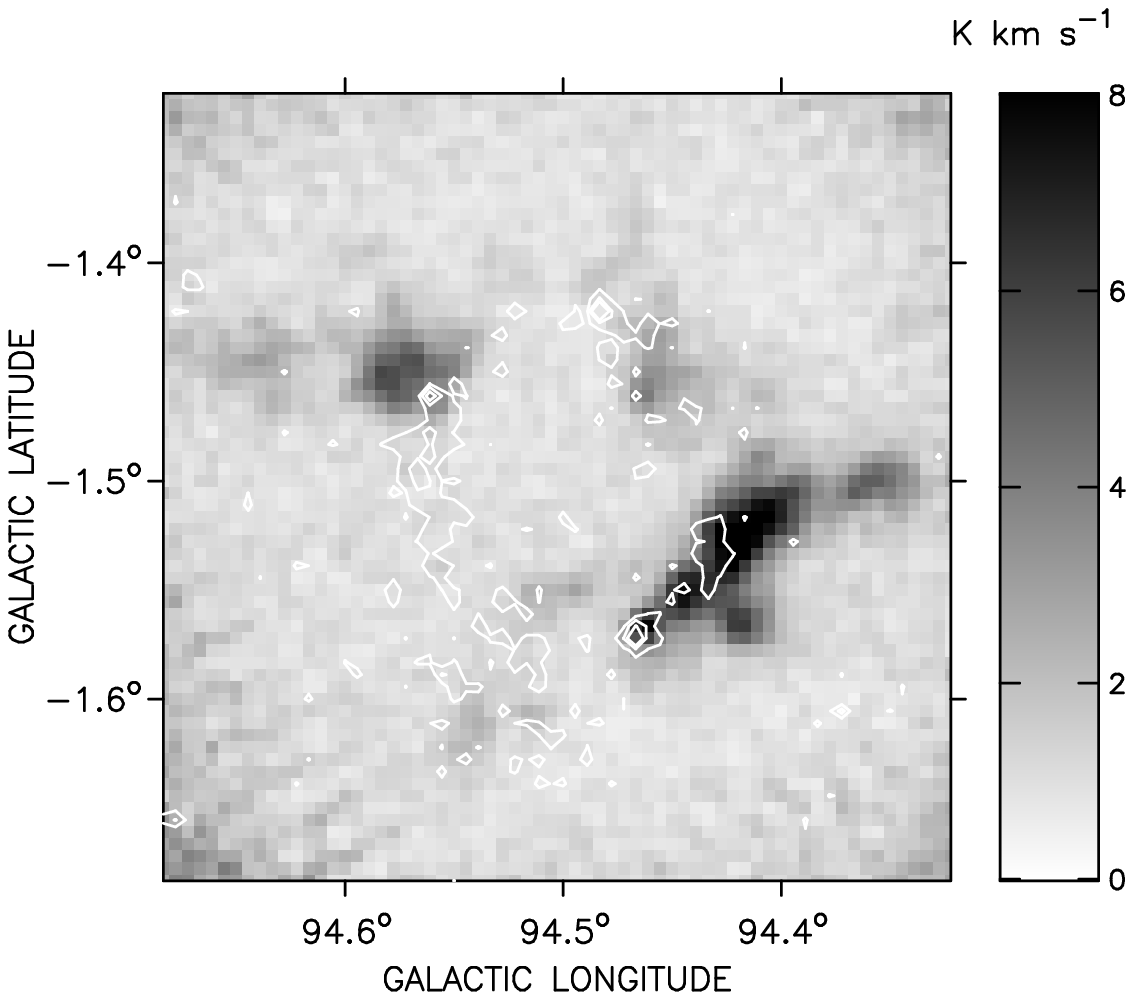}
%\plottwo{f5a.eps}{f5b.eps}
\caption{Top: integrated $^{12}$CO emission around KR~7. Bottom: integrated $^{13}$CO emission around KR~7. The white contours correspond to \emph{MSX} brightness levels of $1.3 \times 10^{-6}$, $2 \times 10^{-6}$, and $3 \times 10^{-6}$ W~m$^{-2}$~sr$^{-1}$ from Figure~\ref{figure-kr7-msxwithradiocontours}.}
\label{figure-kr7-12and13cowithmsxcontours}
\end{figure}

\begin{figure}
\plotone{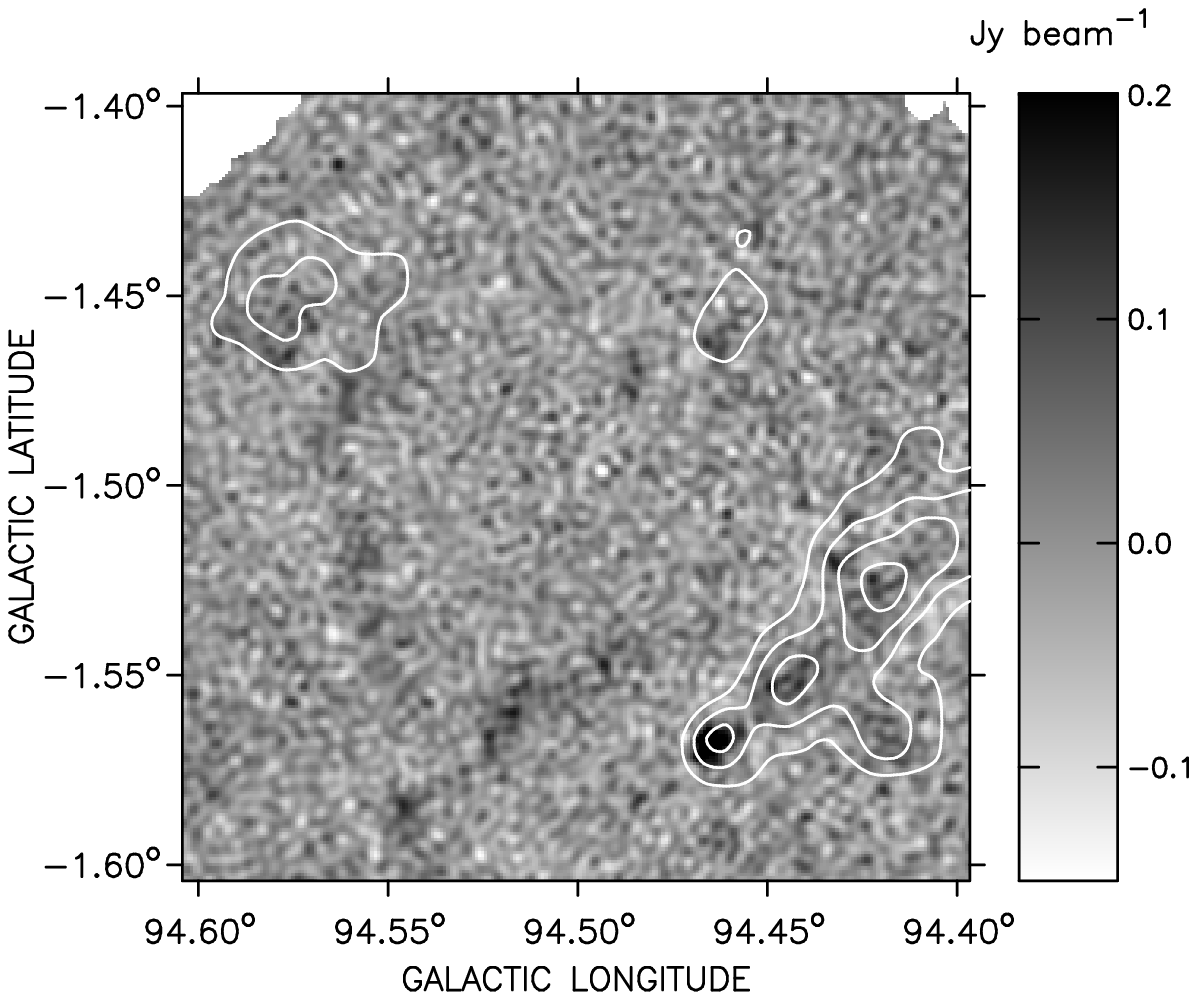}
\caption{Close up view of KR~7 in 850 $\mu$m. The white contours correspond to integrated $^{13}$CO brightness levels of 3, 5, 7, and 9 K~km~s$^{-1}$.}
\label{figure-kr7-submmwith13cocontours}
\end{figure}

For KR~7, the final 850 $\mu$m image is shown in
Figure~\ref{figure-kr7-finalsubmmimage}, and the associated CO emission peaks around $-43$ km~s$^{-1}$ and is mainly found to the Galactic west (smaller Galactic longitude) of the \ion{H}{2} region, see Figure~\ref{figure-kr7-12and13cowithmsxcontours}. There is also a smaller CO concentration to the Galactic northeast of the \ion{H}{2} region. The left panel of Figure~\ref{figure-kr7-12and13cowithmsxcontours} shows integrated $^{12}$CO emission (integrated between $-47.5 < V_{\mathrm{LSR}} < -39.5$ km~s$^{-1}$) and the right panel shows $^{13}$CO (integrated between $-47.5 < V_{\mathrm{LSR}} < -42.8$ km~s$^{-1}$) overlaid with \emph{MSX} A-band contours corresponding to the feature seen in Figure~\ref{figure-kr7-msxwithradiocontours}. For this region and velocity range, C$^{18}$O barely shows a signal above the noise, even when integrated over a velocity range. C$^{18}$O analysis is thus not carried out for KR~7. The $^{12}$CO and $^{13}$CO emission exhibit peaks in the same locations, most notably in a small complex stretching from the location of the sub-mm source up toward the Galactic northwest. This can easily be seen in the lower right corner of Figure~\ref{figure-kr7-submmwith13cocontours} that shows the 850 $\mu$m images overlaid with $^{13}$CO contours. The peak integrated $^{13}$CO emission ($I_{\mathrm{peak}}$) at the location of the sub-mm source has a value of $7.9 \pm 1.2$ K~km~s$^{-1}$, which corresponds to a $^{13}$CO-based $N(\mathrm{H_2}) = (5.7 \pm 0.9) \times 10^{21}$ cm$^{-2}$ or $N(M) = 0.026 \pm 0.004$ g~cm$^{-2}$. The $^{13}$CO and C$^{18}$O (when available) column densities were calculated using the expression \citep[Equation (14.40)]{tools}
\begin{equation}{\label{eq:13co}}
N(\mathrm{CO}) = 1.3 \times 10^{15} \int T_{\mathrm{MB}} \mathrm{d} V ~ \mathrm{cm}^{-2}
\end{equation}
for an assumed main beam temperature of 20 K. Using conversion factors from \citet{simon01} and references therein ($R(^{12}$CO/$^{13}$CO$)=45$ and $R(^{12}$CO/H$_2) = 8 \times 10^{-5}$) the peak H$_2$ column density ($N(\mathrm{H_2})_{\mathrm{peak}}$) was calculated in cm$^{-2}$. The peak mass column density was obtained by multiplying $N(\mathrm{H_2})_{\mathrm{peak}}$ with the H$_2$ molecular mass and a factor of 1.36 to account for elements heavier than hydrogen \citep{simon01}. The uncertainties shown are due to noise in the maps and the $\sim 15 \%$ flux scale uncertainty. They do not include the dominant systematic uncertainty arising from various assumptions in the analysis method, and consequently the values calculated from CO data are expected to be lower limits and accurate within a factor of a few \citep{simon01}. To relate $N(\mathrm{H_2})$ and the integrated $^{12}$CO emission, the conversion factor of $X = 2.3 \times 10^{20}$~cm$^{-2}$~(K~km~s$^{-1}$)$^{-1}$ was used \citep{tools}. By summing over the entire $^{12}$CO feature seen in Figure~\ref{figure-kr7-12and13cowithmsxcontours}, the mass of the molecular material surrounding KR~7 is estimated to be $\sim 3600$ $M_{\sun}$.

The visually dominant KR~7 sub-mm source \citep[JCMTSF J213822.5+501908;][]{scubalegacy-2008ApJS..175..277D} has a flux density ($F_{850}$) of $1.1 \pm 0.2$ Jy within an aperture of $25\arcsec$ around the peak. The deconvolved radius is $\sim 24 \arcsec$ and corresponds to a physical diameter of 0.6 pc at a distance of 2.8 kpc. The maximum brightness ($B_{850}$) is 0.5 Jy~beam$^{-1}$ (S/N $\sim13$). Using standard conversion factors for SCUBA 850 $\mu$m observations \citep[$4.60 \times 10^{22}$ cm$^{-2}$~(Jy~beam$^{-1}$)$^{-1}$ and 6.2 $M_{\sun}$~Jy$^{-1}$~kpc$^{-2}$ from Table A.1 in][]{kauffmann2008AA...487..993K} together with an assumed dust temperature $T_d = 20$ K and a distance of $2.8 \pm 0.4$ kpc \citep{kr7and81distance-2003ApJ...598.1005F}, $B_{850}$ corresponds to a dust-derived peak column density of $N(\mathrm{H_2}) = (2.3 \pm 0.5) \times 10^{22}$ cm$^{-2}$ or $N(M) = 0.11 \pm 0.02$ g~cm$^{-2}$. The dust-based column density is greater than the $^{13}$CO-based column density for the same location. This could be a result of the $^{13}$CO becoming optically thick in the densest regions. The mass of the sub-mm source is $51 \pm 19$ $M_{\sun}$, where the listed uncertainty is the combination of the uncertainty in distance and flux density. Dust-based column density and mass estimates are sensitive to the dust temperature. For example, if the dust temperature was instead assumed to be $T_d = 10$ K, the column density would increase to $6.7 \times 10^{22}$ cm$^{-2}$ or $N(M) = 0.31$ g~cm$^{-2}$ and the mass of the sub-mm core would increase to 150 $M_{\sun}$.

For KR~7, the SCUBA Legacy image corresponding to Figure~\ref{figure-kr7-finalsubmmimage} looks very similar, except for a bright part near the edge of the image to the Galactic northeast ($\ell = 94 \fdg 55$, $b=-1\fdg 45$) that is absent in Figure~\ref{figure-kr7-finalsubmmimage}. The SCUBA Legacy catalog contains several sources. The main source (JCMTSF J213822.5+501908) is the most prominent one in the catalog with $\mathrm{S/N}= 13$, $B_{850} = 0.73 \pm 0.06$ Jy~beam$^{-1}$ and $F_{850} = 1.62$ Jy \citep[absolute flux density uncertainty can be as large as $30 \%$; see][]{scubalegacy-2008ApJS..175..277D}. The corresponding dust-derived peak column density is $N(\mathrm{H_2}) = (3.4 \pm 0.7) \times 10^{22}$ cm$^{-2}$ or $N(M) = 0.16 \pm 0.03$ g~cm$^{-2}$ and the corresponding mass is $78 \pm 32$ $M_{\sun}$ which agrees, within the uncertainties, with our derived mass. The SCUBA Legacy catalog also identifies 10 other faint sources (S/N $\lesssim 4.4$) within and around the region indicated with $^{13}$CO contours in the lower right corner of Figure~\ref{figure-kr7-submmwith13cocontours}. The corresponding locations of the sources have about the same S/N in our image (see Figure~\ref{figure-kr7-finalsubmmimage}) but are harder to identify visually. These probable sources are listed in Table~\ref{table-scubalegacy} together with the corresponding peak mass column densities and masses. In addition, the Legacy catalog lists 21 sources associated with the concentration of $^{13}$CO near $\ell = 94 \fdg 57$, $b=-1\fdg 45$. Inspection of the SCUBA Legacy image shows that these sources are all located in a bright noisy gradient near the edge of the image. We therefore believe most of these sources are spurious and, as we are unable to confidently identify those few sources that may be real, we do not consider any of these sources in further analysis. 

\begin{figure}
\plotone{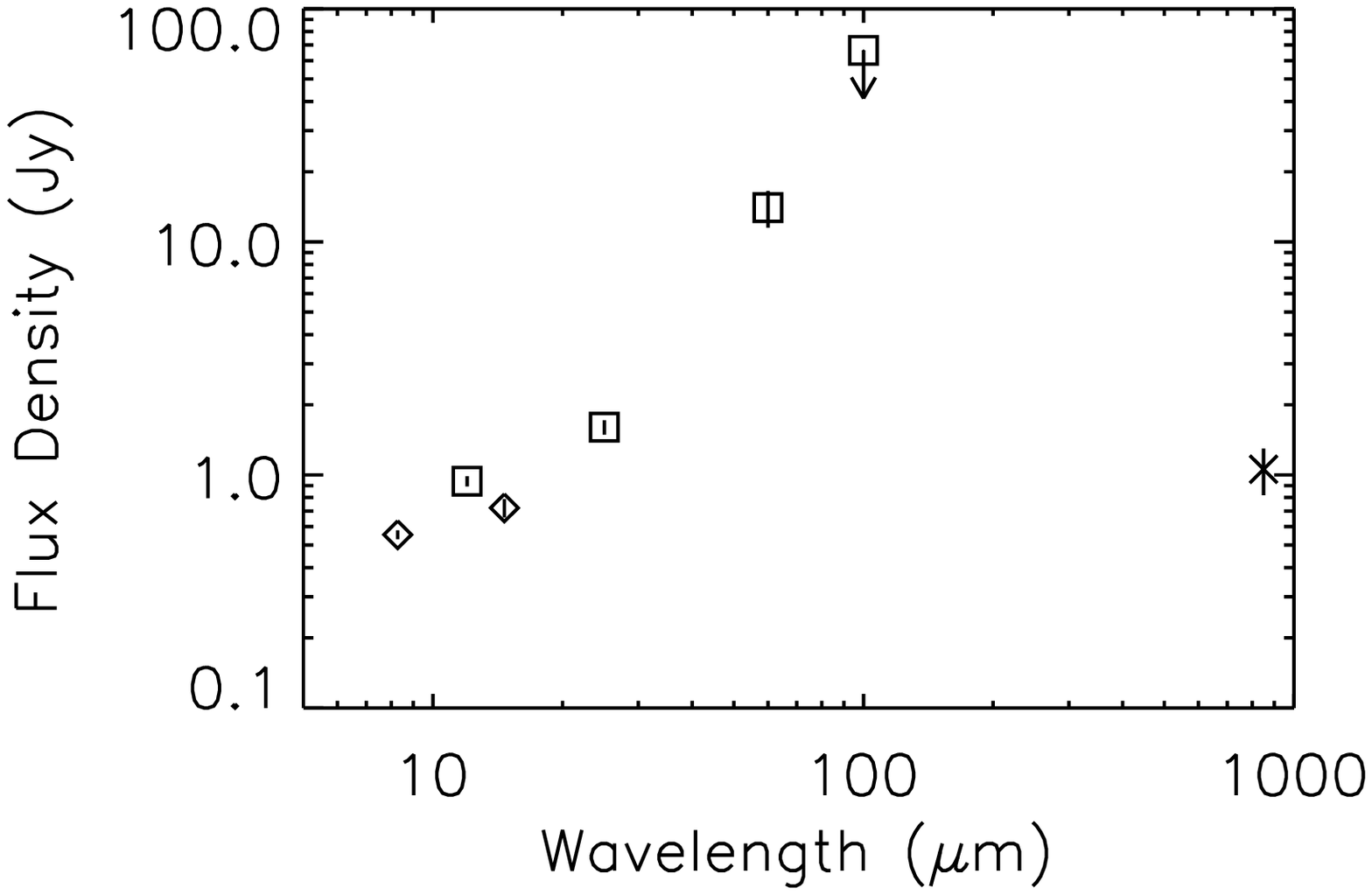}
\caption{Spectral energy distribution for the KR~7 850 $\mu$m source. \emph{IRAS} PSC values are squares, \emph{MSX6C} PSC values are diamonds and the 850 $\mu$m value is the cross. The vertical lines drawn on each symbol indicate $\pm 1\sigma$ uncertainty for each value. The arrow indicates an upper limit.}
\label{figure-kr7-sed}
\end{figure}

The main KR~7 850 $\mu$m source is spatially associated with IRAS 21366+5005 and MSX6C G094.4674-01.5705. In the \emph{IRAS} PSC, 12, 25, and 60 $\mu$m flux densities are listed as being good quality flux densities, while the 100 $\mu$m flux density only is an upper limit. In the \emph{MSX6C} catalog, the source is not detected in the bands C (12.1 $\mu$m, $11.1-13.2$ $\mu$m) and E (21.3 $\mu$m, $18.2-25.1$ $\mu$m), but is detected in the bands A and D (14.7 $\mu$m, $13.5-15.9$ $\mu$m). Figure~\ref{figure-kr7-sed} shows the spectral energy distribution (SED) of this source from 8.3 to 850 $\mu$m. \emph{IRAS} PSC values are squares, \emph{MSX6C} PSC values are diamonds and the 850 $\mu$m cross is the $1.1 \pm 0.2$ Jy found in this work. The arrow indicates an upper limit. Note that in the logarithmic flux density scale, the $1\sigma$ uncertainty lines drawn often are smaller than the printed symbol. Integrating under the curve using the \citet{emerson} method for the \emph{IRAS} fluxes yields an upper limit to the total integrated luminosity ($L_{\mathrm{IR}}$) of $\sim 500$ $L_{\sun}$. The measured and calculated quantities are given in Table~\ref{table-results}.

\begin{figure}
\plotone{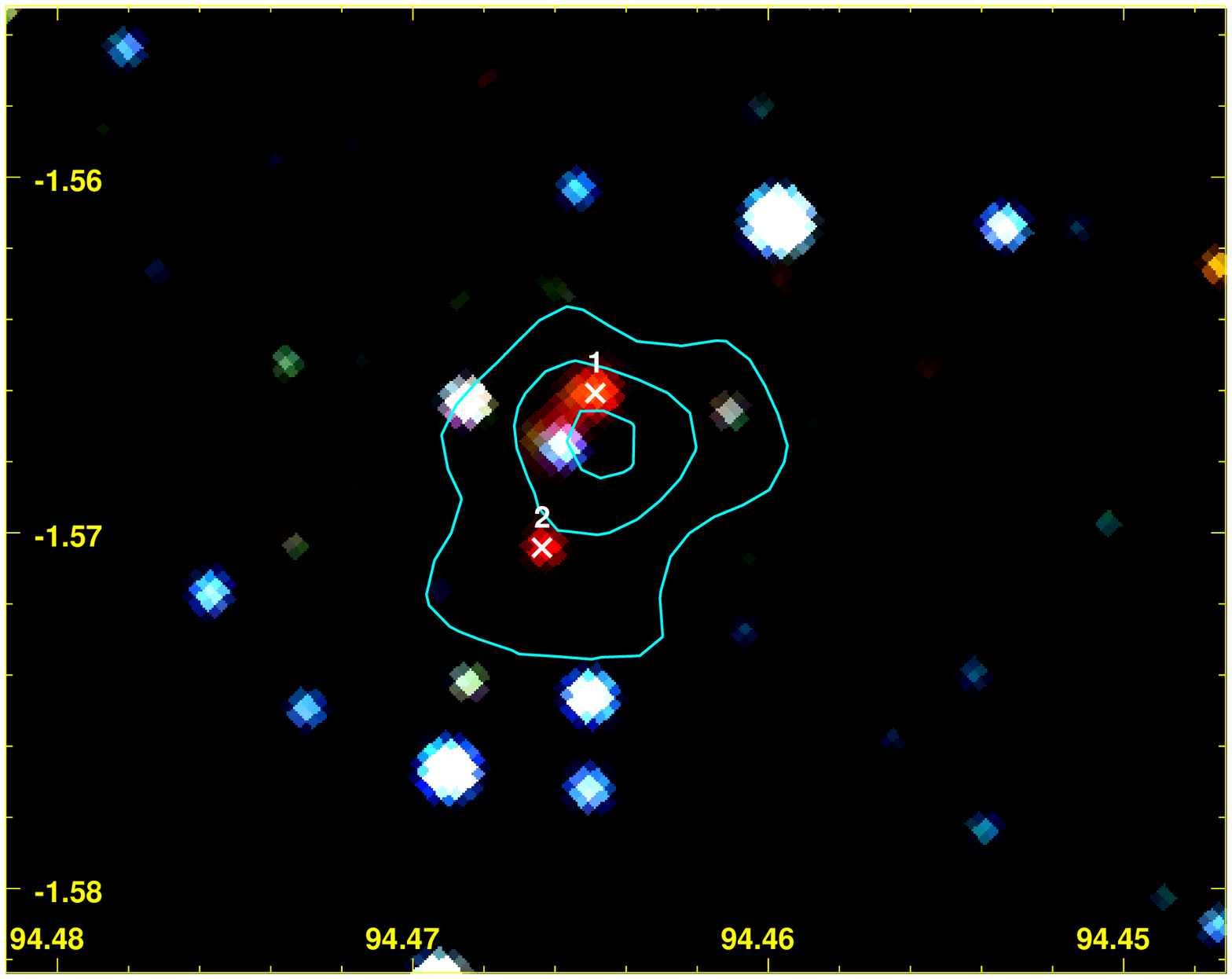}
\caption{$\sim 1\farcm6 \times 2 \arcmin$ RGB composite of 2MASS $J$-band (Blue), $H$-band (Green) and $K_{\mathrm{s}}$-band (Red) images of KR~7. The contours correspond to 850 $\mu$m brightness levels of 0.15, 0.3, and 0.45 Jy~beam$^{-1}$. The crosses are the two possible embedded YSOs, labeled ``1'' and ``2'' in Figure~\ref{figure-kr7-2mass-colorcolor}.}
\label{figure-kr7-2massrgb}
\end{figure}

\begin{figure}
\epsscale{1.1}
\plotone{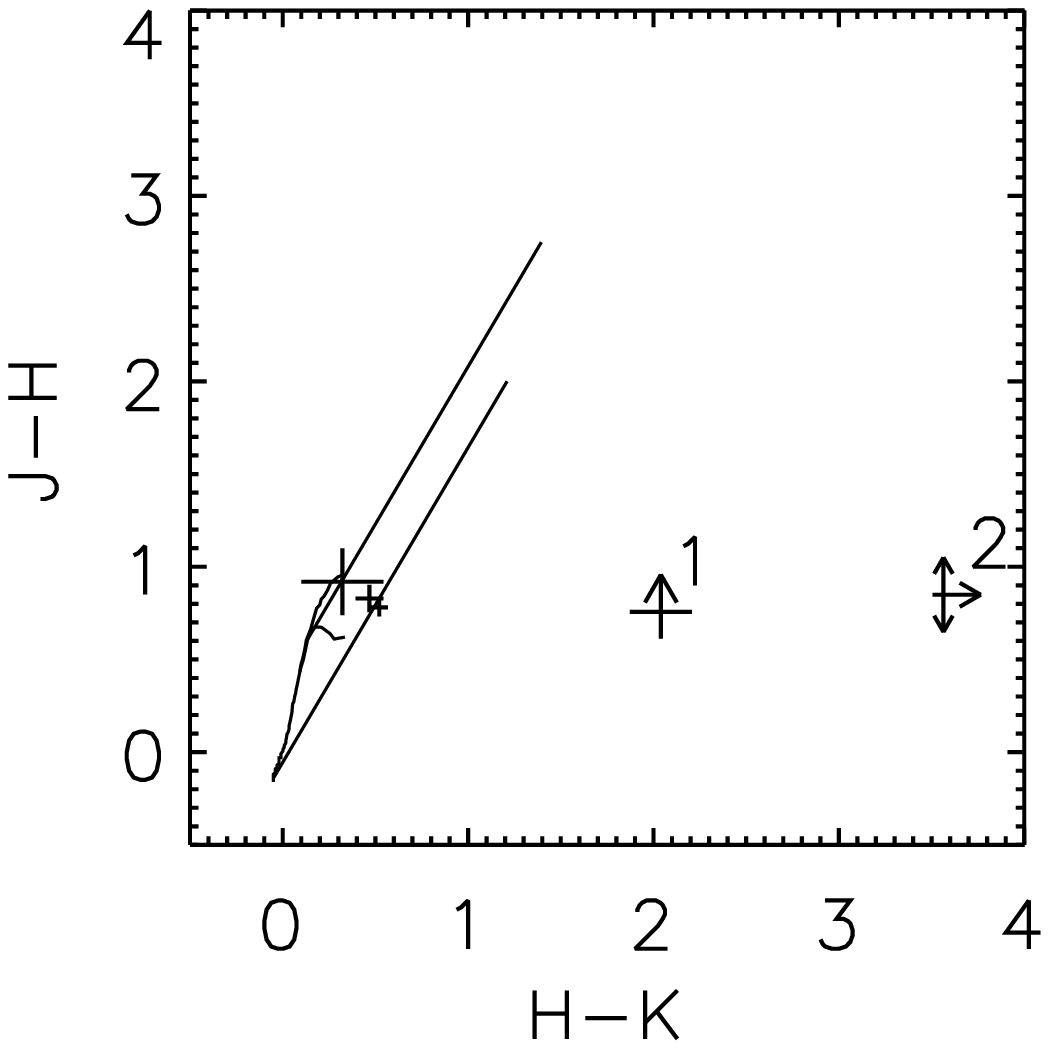}
\plotone{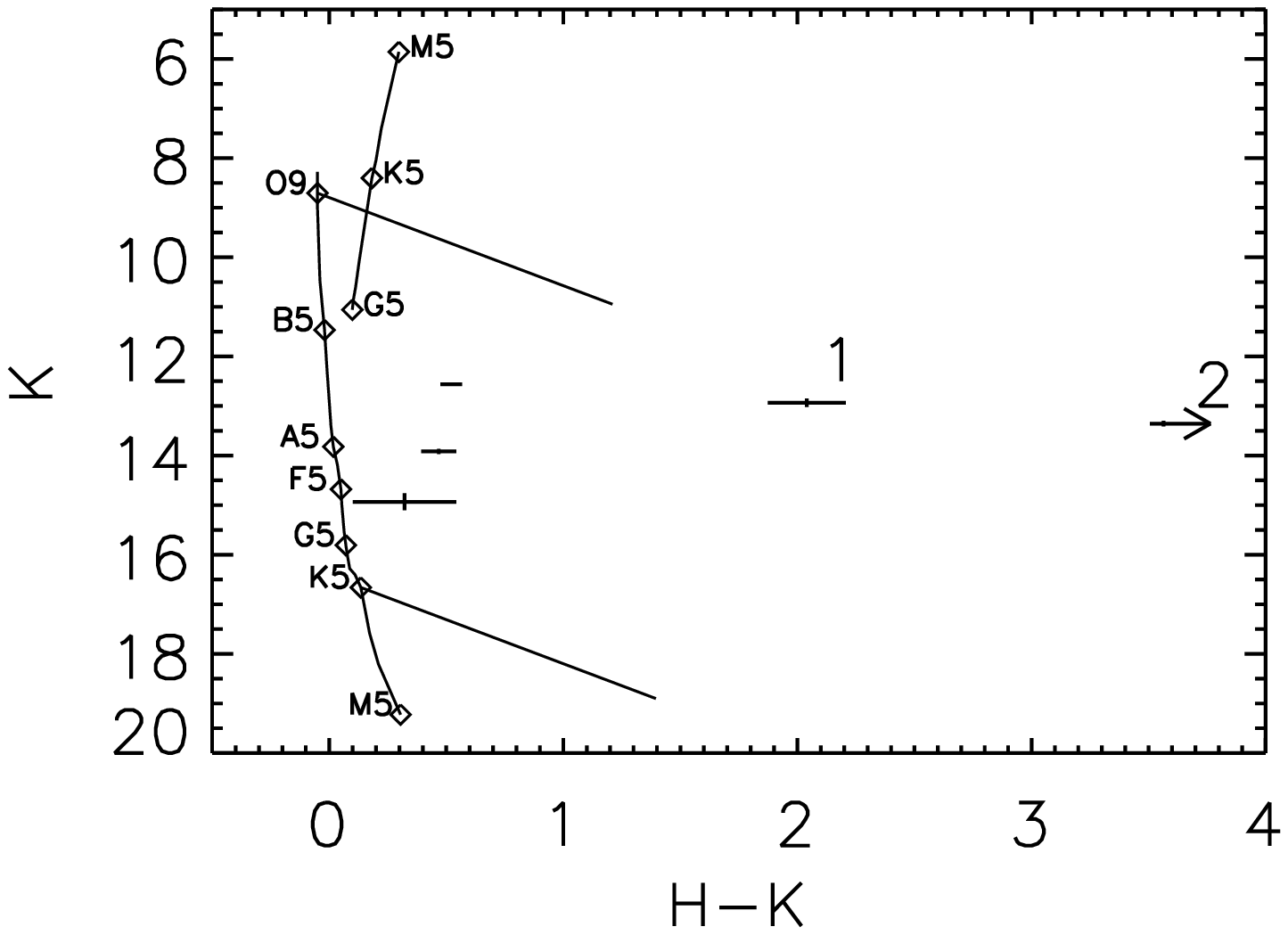}
\caption{Top: color-color diagram for the 2MASS sources within a radius of $0\farcm25$ of the KR~7 850 $\mu$m source. The solid lines in the lower left corner show the intrinsic colors of main-sequence (V) and giant (III) stars. The parallel lines show $A_V = 20$ reddening vectors for a K5~V and a O9~V star. Bottom: color-magnitude diagram for the same 2MASS sources at a distance of 2.8 kpc. The main-sequence and giant branch are shown with some representative spectral types. The parallel lines are again $A_V = 20$ reddening vectors. The size of the plotted crosses for the 2MASS sources indicates the uncertainty in 2MASS photometry. An arrow indicates sources where the 2MASS catalog only lists an upper brightness limit in the relevant band. The three sources near the main-sequence lines are probably foreground stars. The two sources labeled ``1'' and ``2'' are the two possible embedded YSOs.}
\label{figure-kr7-2mass-colorcolor}
\epsscale{1.0}
\end{figure}

Figure~\ref{figure-kr7-2massrgb} shows a composite image of the region ($\sim 1\farcm6  \times 2 \arcmin$) surrounding the dominant KR~7 850 $\mu$m source, made from Two Micron All Sky Survey (2MASS) $J$-band (Blue), $H$-band (Green) and $K_{\mathrm{s}}$-band (Red) images. The contours, 850 $\mu$m brightness levels of 0.15, 0.3, and 0.45 Jy~beam$^{-1}$, trace the very central part of the sub-mm source. The 2MASS catalog lists seven sources within a radius of $\sim 0\farcm25$ from the location of the peak $B_{850}$, an area roughly corresponding to the extent of the 850 $\mu$m 0.15 Jy~beam$^{-1}$ contour. Figure~\ref{figure-kr7-2mass-colorcolor} shows these sources plotted in a 2MASS color-color diagram (top) and a color-magnitude diagram (bottom). The solid lines in the lower left corner of the color-color diagram show the intrinsic colors of main-sequence (V) and giant (III) stars \citep{koornneef-1983A&A...128...84K,bessellandbrett-1988PASP..100.1134B}. The parallel lines show $A_V = 20$ reddening vectors for a K5~V and a O9~V star derived using the infrared extinction law of \citet{riekeandlebofsky-1985ApJ...288..618R}. For the color-magnitude diagram, the main-sequence and giant branch are shown with some representative spectral types at a distance of 2.8 kpc. The parallel lines are again $A_V = 20$ reddening vectors for a K5~V and a O9~V star. The size of the plotted crosses for the 2MASS sources in both diagrams indicates the uncertainty in 2MASS photometry. An arrow indicates sources where the 2MASS catalog only lists an upper brightness limit in the relevant band. The three sources near the main-sequence lines are probably foreground stars. The two sources labeled ``1'' and ``2'' in Figure~\ref{figure-kr7-2massrgb} and Figure~\ref{figure-kr7-2mass-colorcolor} have $H - K \geq 2$ and are probably highly embedded YSOs that would be associated with the cold dust core detected in the 850 $\mu$m image.

\subsection{KR~81}\label{subsec-kr81analysis}
KR~81 (Sh 2-165) is a $6 \arcmin$ diameter 1420 MHz radio continuum source with a flux density $F_{1420} = 607 \pm 40$ mJy \citep{kerton-krcatalogrevisted-2006MNRAS.373.1203K}. The estimated flux of ionizing photons, $\log{N_L} = 47.2$ corresponds to a single B0~V star \citep{crowther2005IAUS..227..389C}. This agrees well with observations of the exciting star BD $+61$ $2494$, listed as a B0~V star in \citet{crampton-fisher1974PDAO...14..283C} or B0.5~V in \citet{hunterandmassey-1990AJ.....99..846H}.

\begin{figure}
\plotone{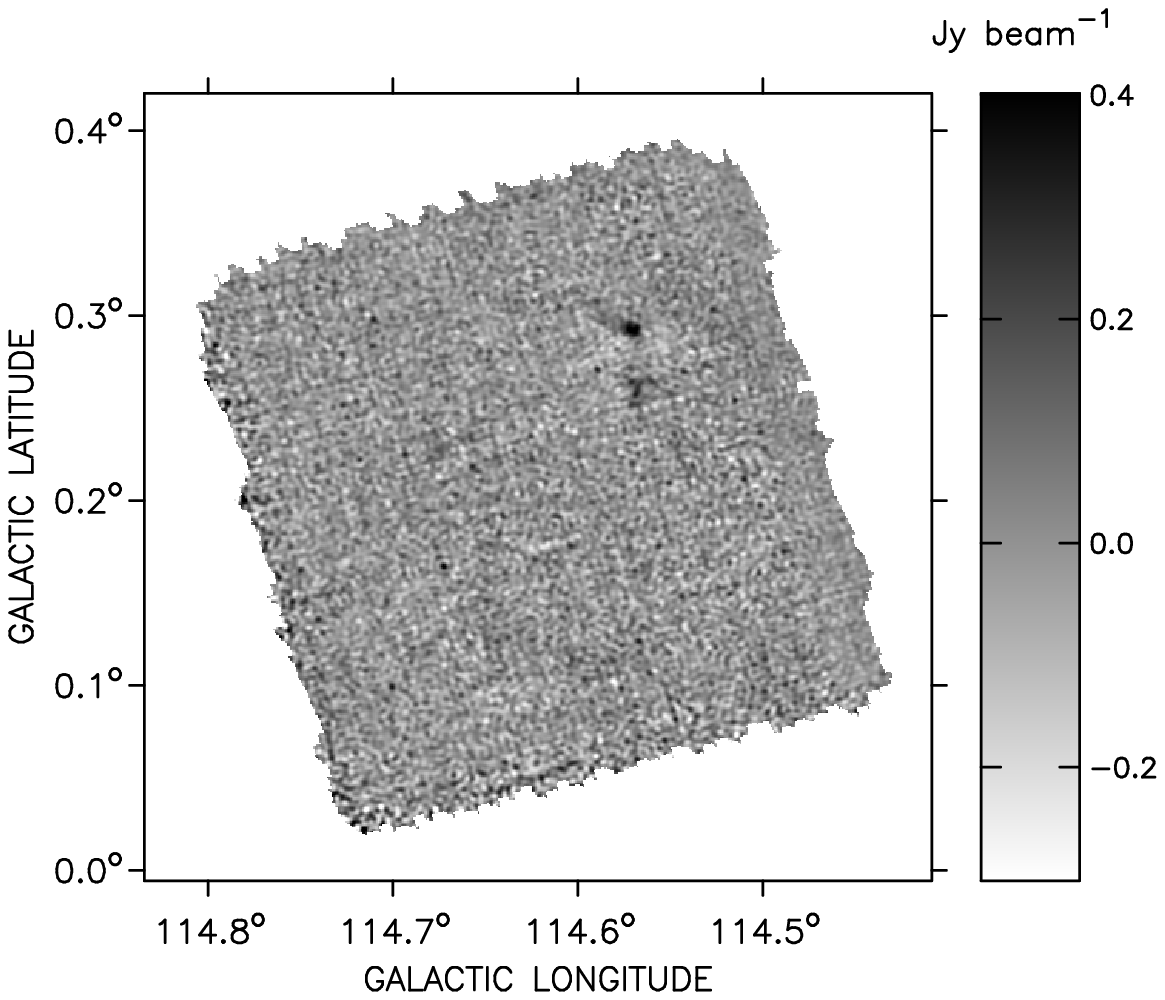}
\caption{Final 850 $\mu$m image of KR~81.}
\label{figure-kr81-finalsubmmimage}
\end{figure}

\begin{figure}
\epsscale{1.4}
\plotone{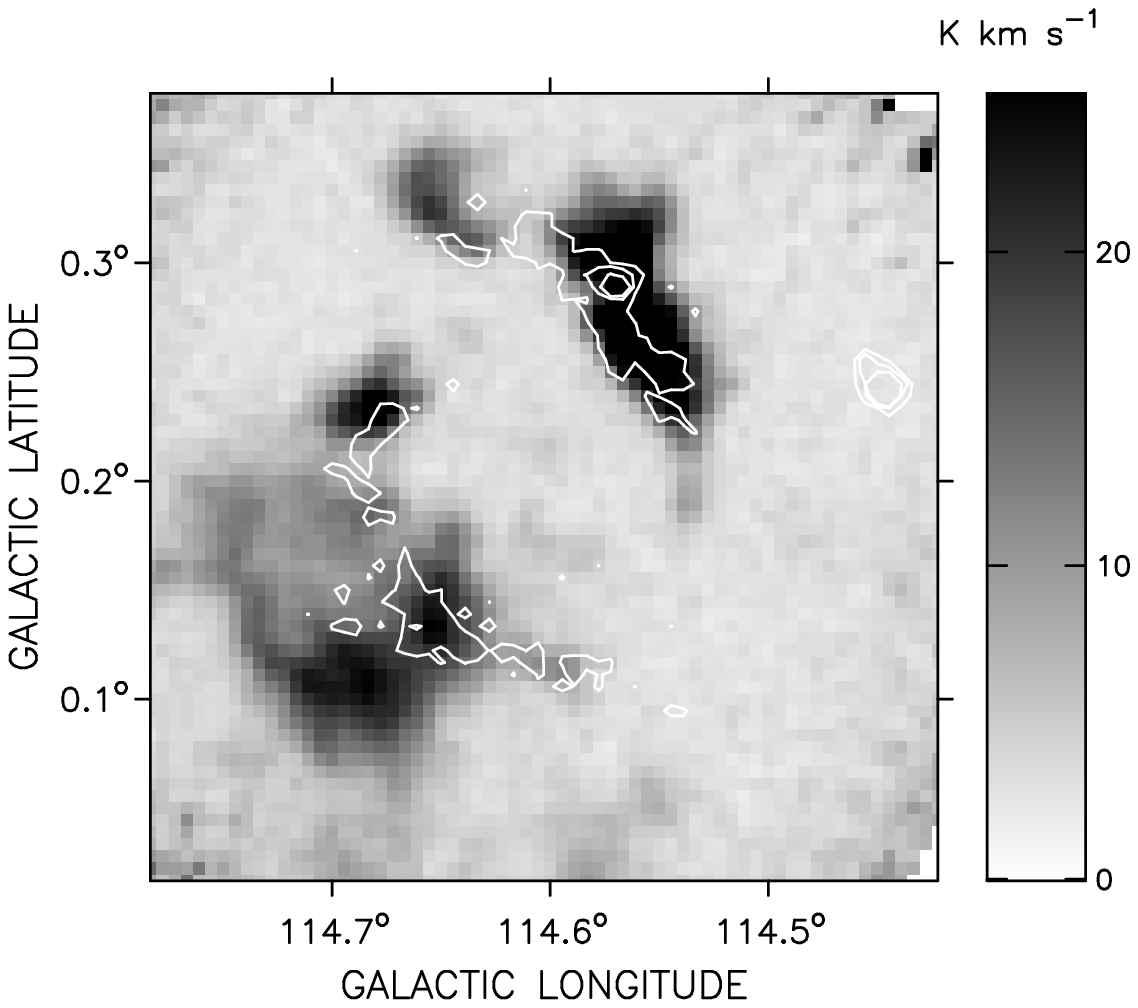}
\plotone{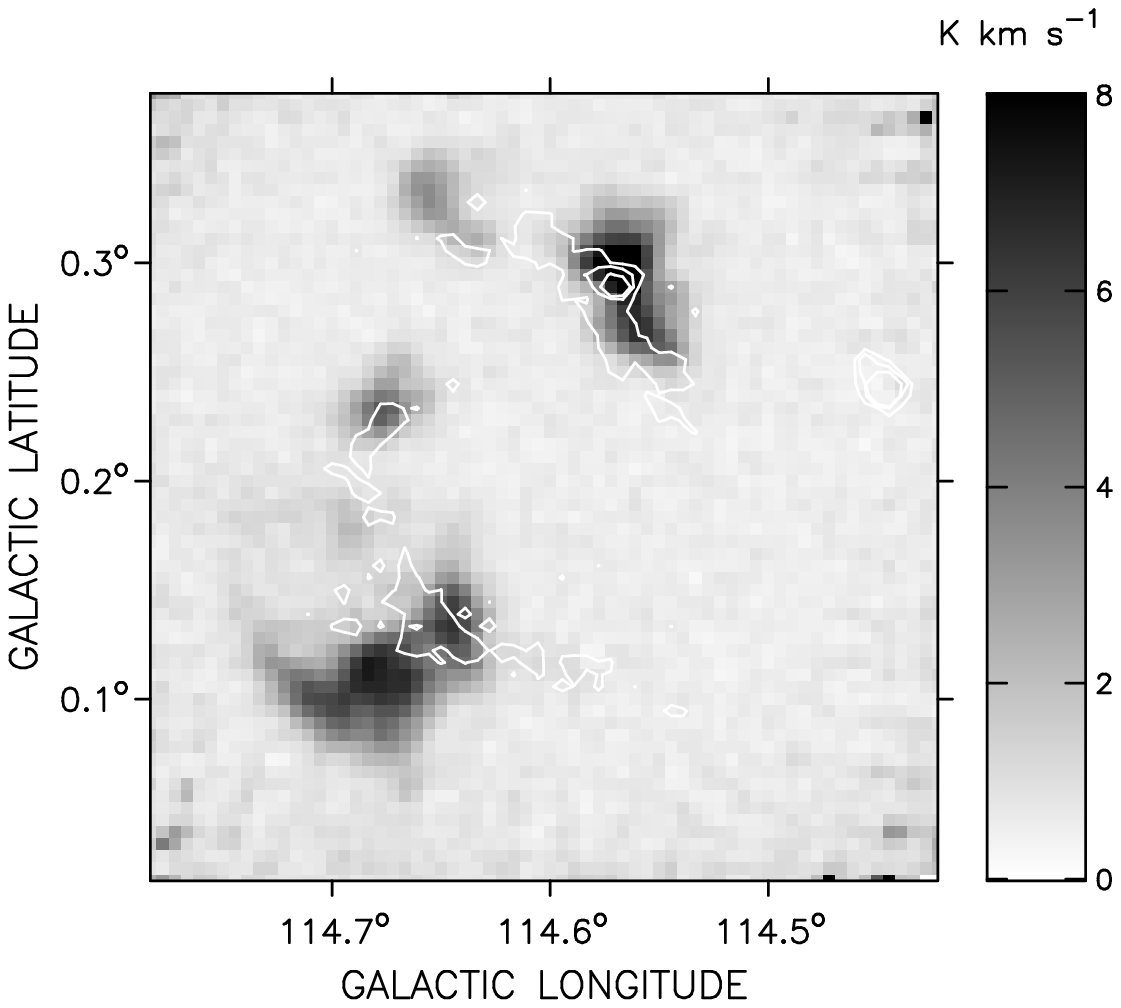}
%\plottwo{f11a.eps}{f11b.eps}
\caption{Top: integrated $^{12}$CO emission around KR~81. Bottom: integrated $^{13}$CO emission around KR~81. The white contours correspond to \emph{MSX} brightness levels of $1.5 \times 10^{-6}$, $3 \times 10^{-6}$, and $6 \times 10^{-6}$ W~m$^{-2}$~sr$^{-1}$ from Figure~\ref{figure-kr81-msxwithradiocontours}.}
\label{figure-kr81-12and13cowithmsxcontours}
\end{figure}

\begin{figure}
\plotone{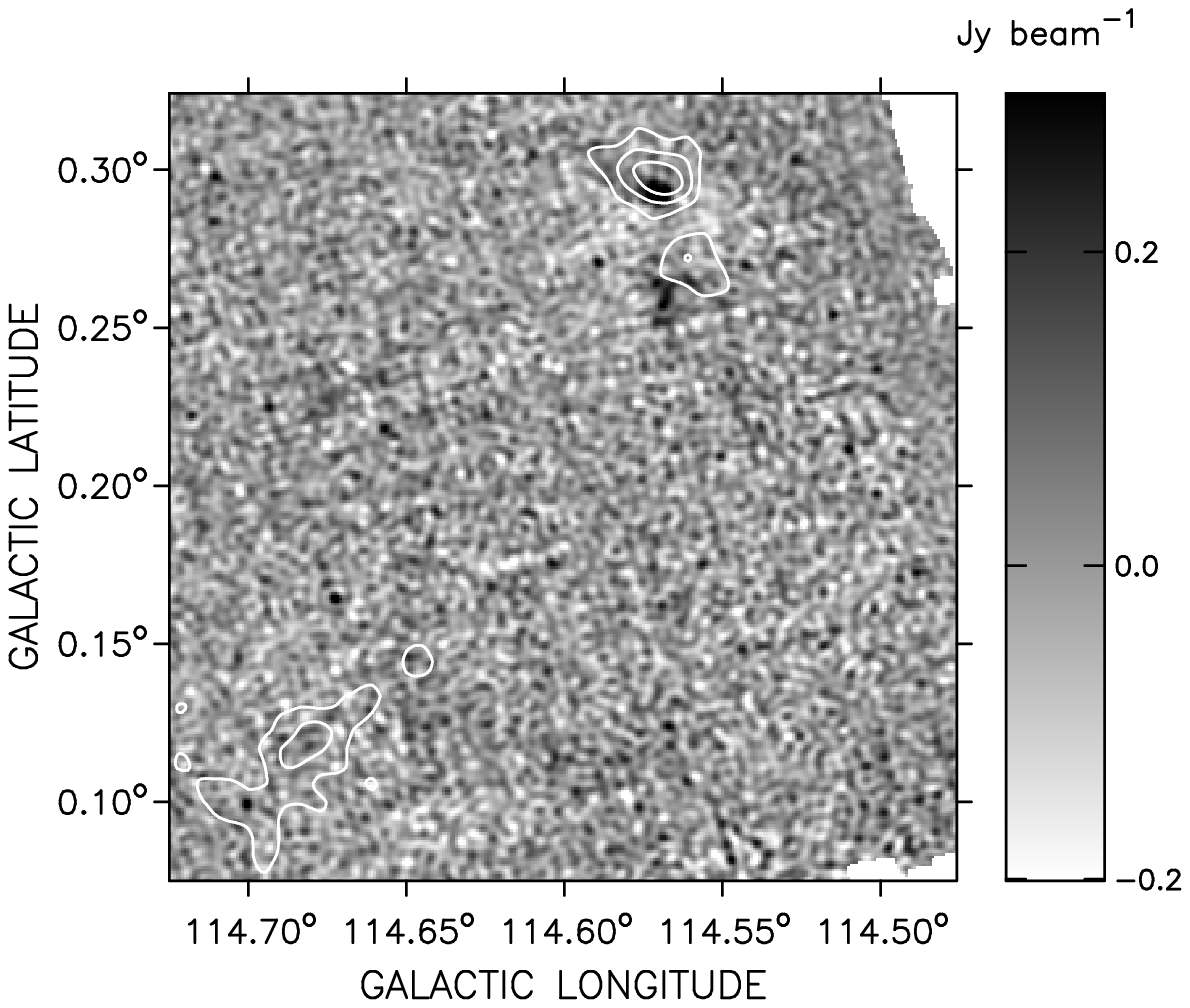}
\caption{Close up view of KR~81 in 850 $\mu$m. The white contours correspond to integrated C$^{18}$O brightness levels of 0.5, 0.75, and 1.0 K~km~s$^{-1}$.}
\label{figure-kr81-submmwithc18ocontours}
\end{figure}

For KR~81, the final 850 $\mu$m image is shown in
Figure~\ref{figure-kr81-finalsubmmimage}, and the associated CO emission peaks around $-34$ km~s$^{-1}$ and is mainly found to the Galactic northwest (smaller Galactic longitude) of the \ion{H}{2} region, where the prominent sub-mm source, discussed in the next paragraph, is located (see Figure~\ref{figure-kr81-12and13cowithmsxcontours}). There is also CO to the Galactic southeast of the \ion{H}{2} region. Figure~\ref{figure-kr81-12and13cowithmsxcontours} shows integrated CO emission from $^{12}$CO (left panel, integrated between $-37.5 < V_{\mathrm{LSR}} < -29.4$ km~s$^{-1}$) and $^{13}$CO (right panel, integrated between $-35.5 < V_{\mathrm{LSR}} < -31.8$ km~s$^{-1}$) overlaid with \emph{MSX} A-band contours corresponding to the feature seen in Figure~\ref{figure-kr81-msxwithradiocontours}. The $I_{\mathrm{peak}}$ of the integrated $^{13}$CO emission at the location of the sub-mm source has a value of $13.5 \pm 2.0$ K~km~s$^{-1}$, which corresponds to a $^{13}$CO-based $N(\mathrm{H_2}) = (9.8 \pm 1.5) \times 10^{21}$ cm$^{-2}$ or $N(M) = 0.045 \pm 0.007$ g~cm$^{-2}$. C$^{18}$O emission is detected toward this region (see Figure~\ref{figure-kr81-submmwithc18ocontours}) and shows a strong peak at the location of the sub-mm source. Using the isotope number ratio R(C$^{18}$O/H$_2) = 1.7 \times 10^{-7}$ \citep{tools}, the C$^{18}$O-based column density is $N(\mathrm{H_2}) = (1.0 \pm 0.2) \times 10^{22}$ cm$^{-2}$ or $N(M) = 0.045 \pm 0.007$ g~cm$^{-2}$, very close to the estimate from $^{13}$CO. By summing over the entire $^{12}$CO feature seen in Figure~\ref{figure-kr81-12and13cowithmsxcontours}, the mass of the molecular material surrounding KR~81 is estimated to be $\sim 1900$ $M_{\sun}$.

The prominent KR~81 sub-mm source (JCMTSF J233917.8+615914) has a flux density of $1.2 \pm 0.3$ Jy within an aperture of $25\arcsec$ around the peak. The deconvolved radius is $\sim 24 \arcsec$ and corresponds to a physical diameter of 0.4 pc at a distance of $1.9 \pm 0.4$ kpc. $B_{850}$ is 0.5 Jy~beam$^{-1}$ (S/N of $\sim6.7$). This corresponds to a dust-derived peak column density of $N(\mathrm{H_2}) = (2.2 \pm 0.5) \times 10^{22}$ cm$^{-2}$ or $N(M) = 0.10 \pm 0.03$ g~cm$^{-2}$. This dust-based column density is greater than the CO-based column density for the same location, again a possible opacity effect. The mass of the sub-mm source is $27 \pm 13$ $M_{\sun}$. For $T_d = 10$ K, the column density would increase to $6.4 \times 10^{22}$ cm$^{-2}$ or $N(M) = 0.30$ g~cm$^{-2}$ and the mass would increase to 80 $M_{\sun}$.

The SCUBA Legacy image for KR~81 looks very similar to Figure~\ref{figure-kr81-finalsubmmimage}. The main source (JCMTSF J233917.8+615914) is the most prominent one in the catalog with $\mathrm{S/N}= 13.8$, $F_{850} =2.17$ Jy and $B_{850} = 0.84 \pm 0.06$ Jy~beam$^{-1}$. The corresponding dust-derived peak column density is $N(\mathrm{H_2}) = (3.9 \pm 0.8) \times 10^{22}$ cm$^{-2}$ or $N(M) = 0.18 \pm 0.04$ g~cm$^{-2}$ and the corresponding mass is $48 \pm 25$ $M_{\sun}$. This mass is larger than determined in the previous paragraph because of the larger flux, stemming from to the larger source area and peak value in the Legacy catalog. The SCUBA Legacy catalog contains three additional sources near the main source. The sources are listed in Table~\ref{table-scubalegacy} together with the corresponding peak mass column densities and masses. Five more sources are found near $\ell = 114 \fdg 5$, $b=0\fdg 14$.  However the reality of these sources is very questionable as they all have low listed S/N ($\leq 3.6$) and are located in a region with no detectable CO emission. Therefore we do not include them in our analysis. The SCUBA Legacy catalog has no detected sources around the CO concentration near $\ell = 114 \fdg 68$, $b=0\fdg 12$.

\begin{figure}
\plotone{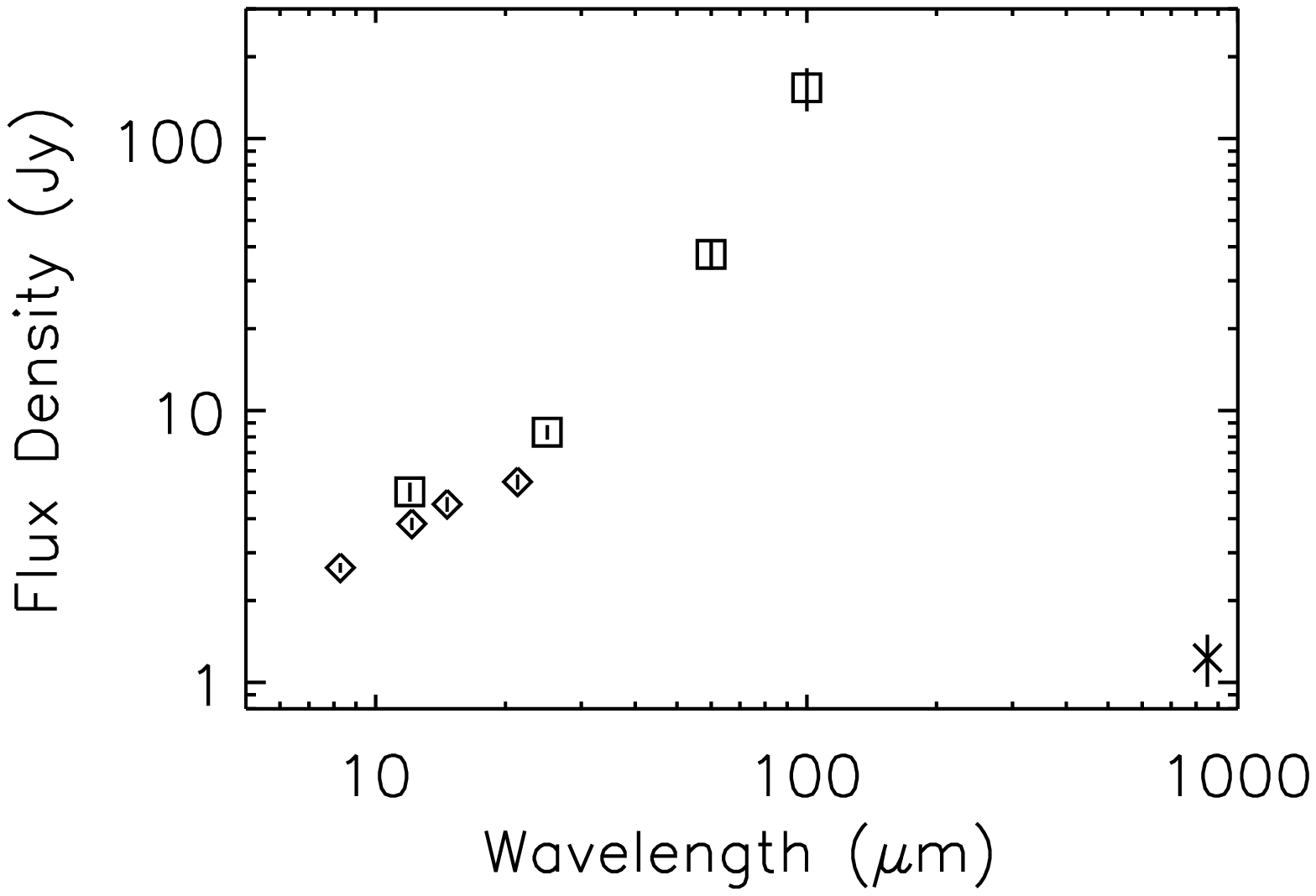}
\caption{Spectral energy distribution for the KR~81 850 $\mu$m source. \emph{IRAS} PSC values are squares, \emph{MSX6C} PSC values are diamonds and the 850 $\mu$m value is the cross. The vertical lines drawn on each symbol indicate $\pm 1\sigma$ uncertainty for each value.}
\label{figure-kr81-sed}
\end{figure}

The main KR~81 850 $\mu$m source is spatially associated with IRAS 23369+6142 and MSX6C G114.5696+00.2899. The \emph{IRAS} PSC entries are all listed as either good or moderate quality, and the source is detected in all \emph{MSX} bands. Figure~\ref{figure-kr81-sed} shows the SED of this source from 8.3 to 850 $\mu$m. Integrating under the curve yields $L_{IR} = 700 \pm 300$ $L_{\sun}$, where the uncertainty is the combination of the uncertainty in distance and \emph{IRAS} flux density. The measured and calculated quantities are given in Table~\ref{table-results}.

\begin{figure}
\plotone{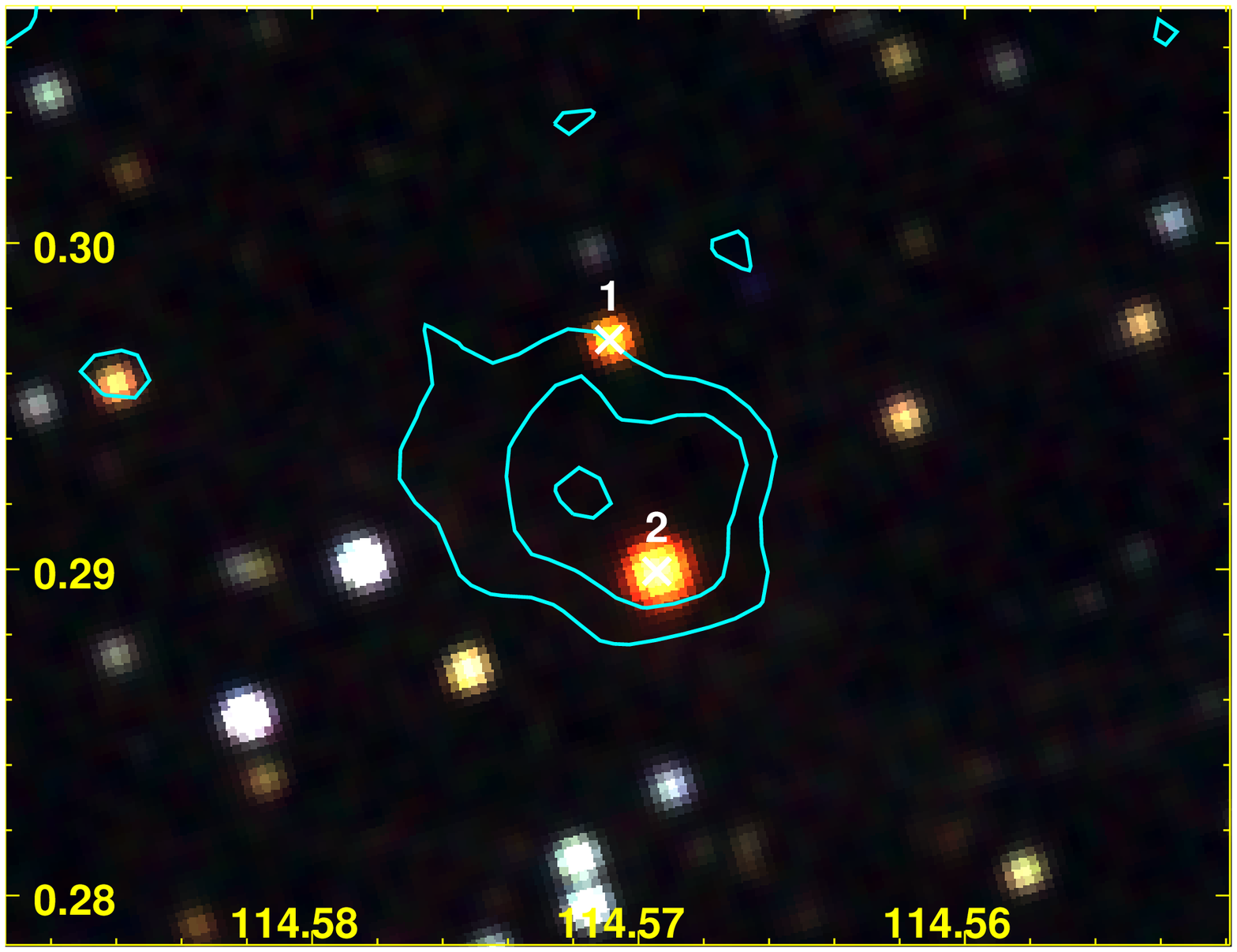}
\caption{$\sim 2 \arcmin \times 2 \arcmin$ RGB composite of 2MASS $J$-band (Blue), $H$-band (Green) and $K_{\mathrm{s}}$-band (Red) images of KR~81. The contours correspond to 850 $\mu$m brightness levels of 0.15, 0.3, and 0.45 Jy~beam$^{-1}$. The crosses are the two possible background giant stars, labeled ``1'' and ``2'' in Figure~\ref{figure-kr81-2mass-colorcolor}.}
\label{figure-kr81-2massrgb}
\end{figure}

\begin{figure}
\epsscale{1.1}
\plotone{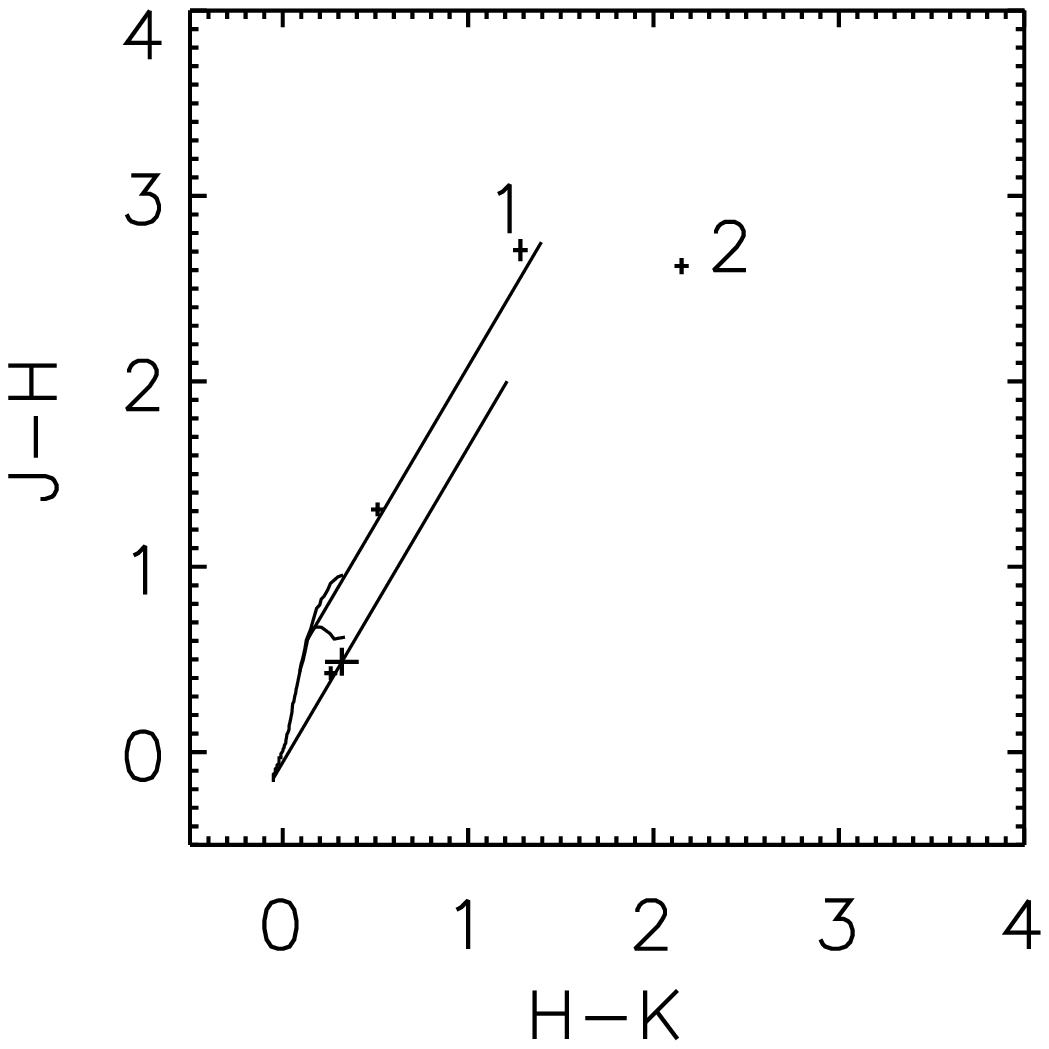}
\plotone{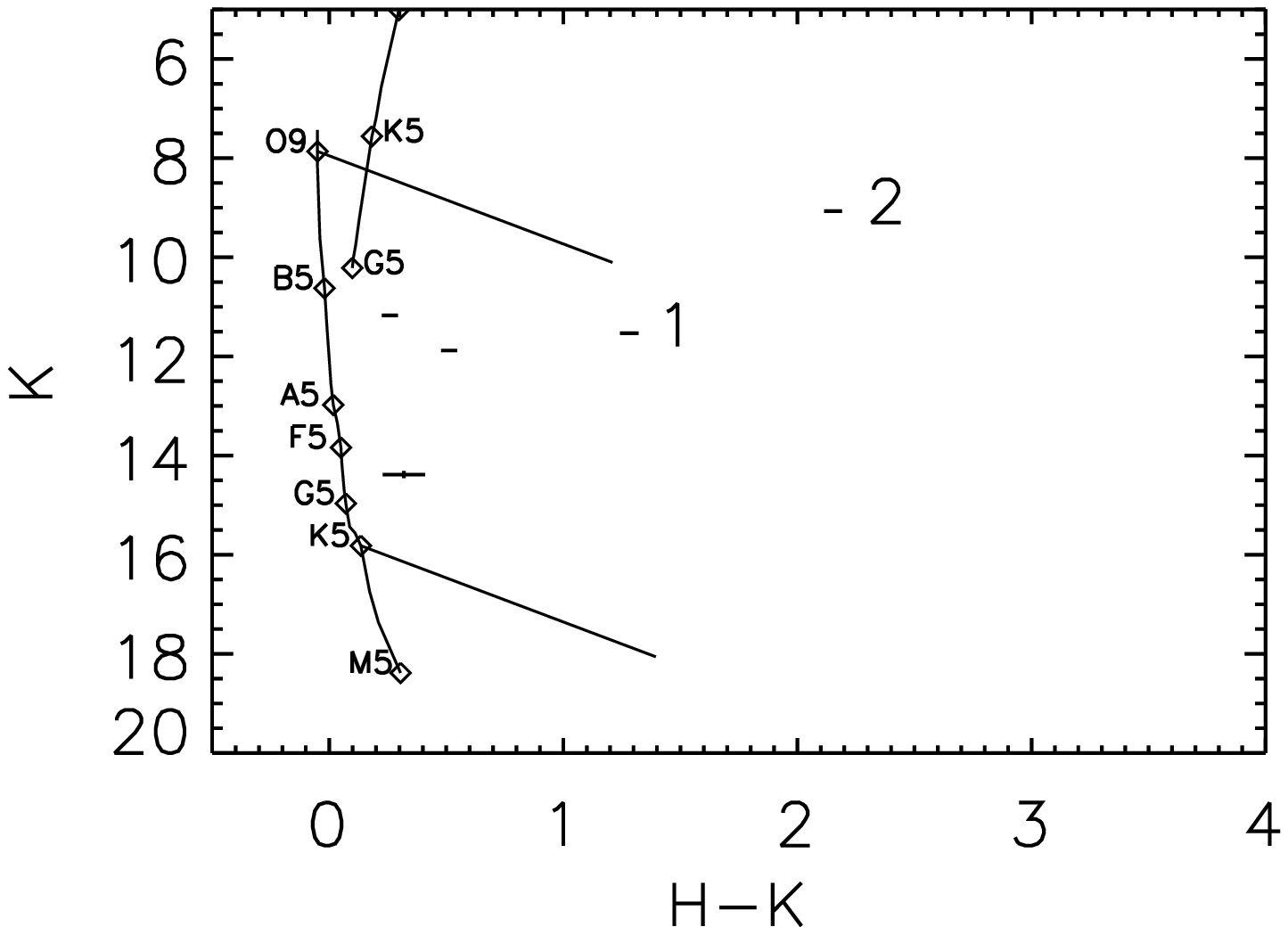}
\caption{Top: color-color diagram for the 2MASS sources within a radius of $0\farcm5$ of the KR~81 850 $\mu$m source. The solid lines in the lower left corner show the intrinsic colors of main-sequence (V) and giant (III) stars. The parallel lines show $A_V = 20$ reddening vectors for a K5~V and a O9~V star. Bottom: color-magnitude diagram for the same 2MASS sources at a distance of 1.9 kpc. The main-sequence and giant branch are shown with some representative spectral types. The parallel lines are again $A_V = 20$ reddening vectors. The size of the plotted crosses for the 2MASS sources indicates the uncertainty in 2MASS photometry. The three sources near the main-sequence lines are probably foreground stars. The two sources labeled ``1'' and ``2'' are the two possible background giant stars.}
\label{figure-kr81-2mass-colorcolor}
\end{figure}

Figure~\ref{figure-kr81-2massrgb} shows a composite image of the region ($\sim 2 \arcmin \times 2 \arcmin$) surrounding the dominant KR~81 850 $\mu$m source, made from 2MASS $J$-band (Blue), $H$-band (Green) and $K_{\mathrm{s}}$-band (Red) images. The contours are 850 $\mu$m brightness levels of 0.15, 0.3, and 0.45 Jy~beam$^{-1}$, and indicate the very central part of the sub-mm source, seen as the darkest small ``blob'' in Figures~\ref{figure-kr81-finalsubmmimage} and \ref{figure-kr81-submmwithc18ocontours}. The 2MASS catalog lists five sources within a radius of $\sim 0\farcm5$ from the location of the peak $B_{850}$, an area roughly corresponding to the extent of the 850 $\mu$m 0.15 Jy~beam$^{-1}$ contour. Figure~\ref{figure-kr81-2mass-colorcolor} shows these sources plotted in a 2MASS color-color diagram (top) and a color-magnitude diagram for a distance of 1.9 kpc (bottom), like Figure~\ref{figure-kr7-2mass-colorcolor}. The three sources nearest to the main-sequence/giant branch lines line are probably foreground stars. The two sources with $J - H \geq 2.5$, marked with crosses in Figure~\ref{figure-kr81-2massrgb}, could be background giant stars. Star number 1 in Figure~\ref{figure-kr81-2mass-colorcolor} is located where a very reddened ($A_V \sim 20$) red giant would be. Star number 2 could be a very reddened Long-Period Variable (LPV) giant star. LPVs are intrinsically very red due to molecular blanketing and cooler continuum temperatures \citep{bessellandbrett-1988PASP..100.1134B}. No obvious embedded YSO candidates are found in the cold dust core detected in the 850 $\mu$m image using 2MASS color-color and color-magnitude diagrams.

\subsection{KR~120}\label{subsec-kr120analysis}
KR~120 (Sh 2-187) is a $6 \arcmin$ diameter 1420 MHz radio continuum source with a flux density $F_{1420} = 928 \pm 28$ mJy \citep{kerton-krcatalogrevisted-2006MNRAS.373.1203K}. The estimated flux of ionizing photons, $\log{N_L} = 47.2$ corresponds to a B0~V star \citep{crowther2005IAUS..227..389C}. \citet{kr120distance-2007A&A...470..161R} suggest that a B2.5~V star could be the exciting star for KR~120, but a cluster of such stars would be needed to match the estimated ionizing photon flux. As with KR~7, the identity of the exciting star of this region remains an open question.

\begin{figure}
\plotone{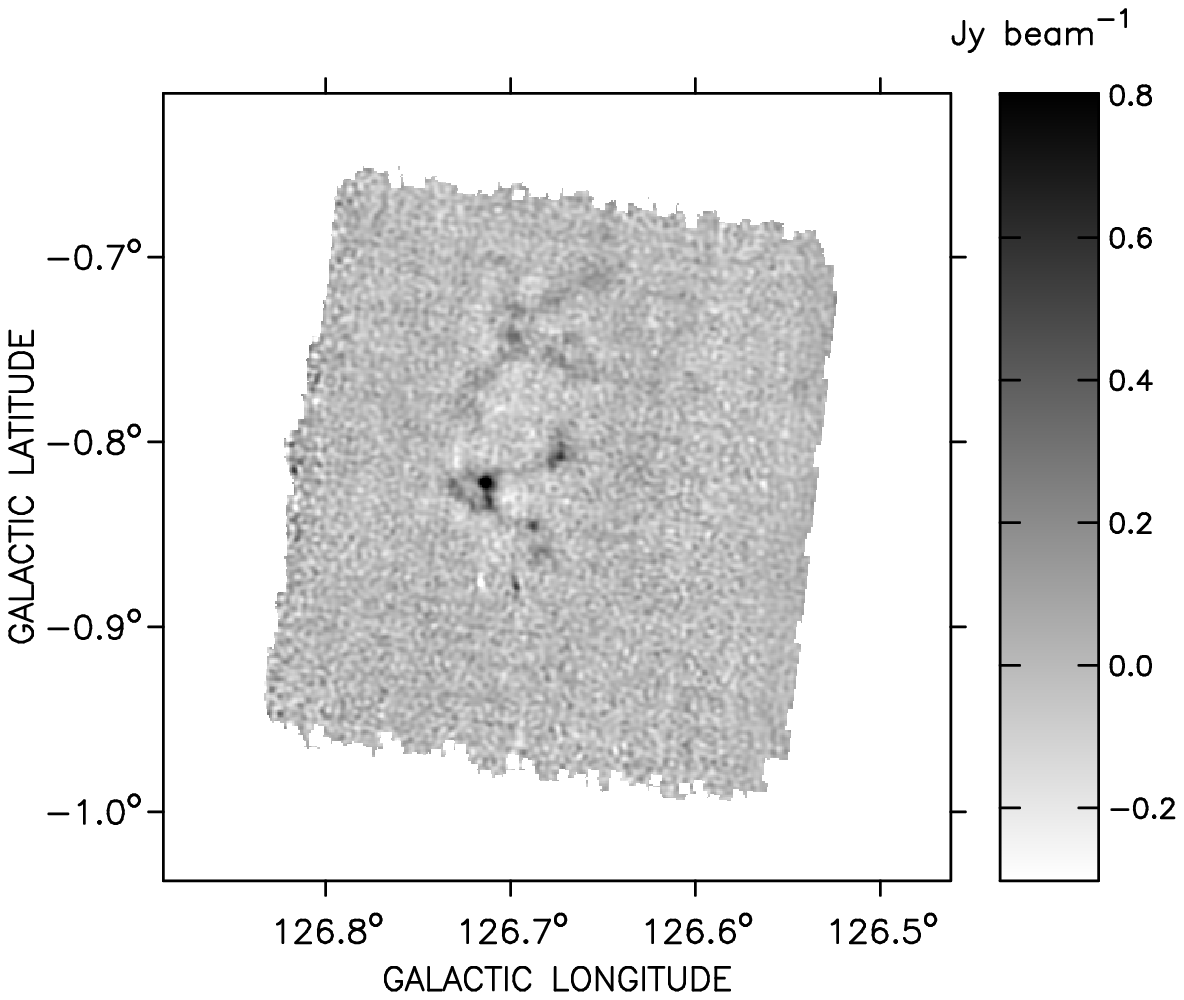}
\caption{Final 850 $\mu$m image of KR~120.}
\label{figure-kr120-finalsubmmimage}
\end{figure}

\begin{figure}
\epsscale{1.4}
\plotone{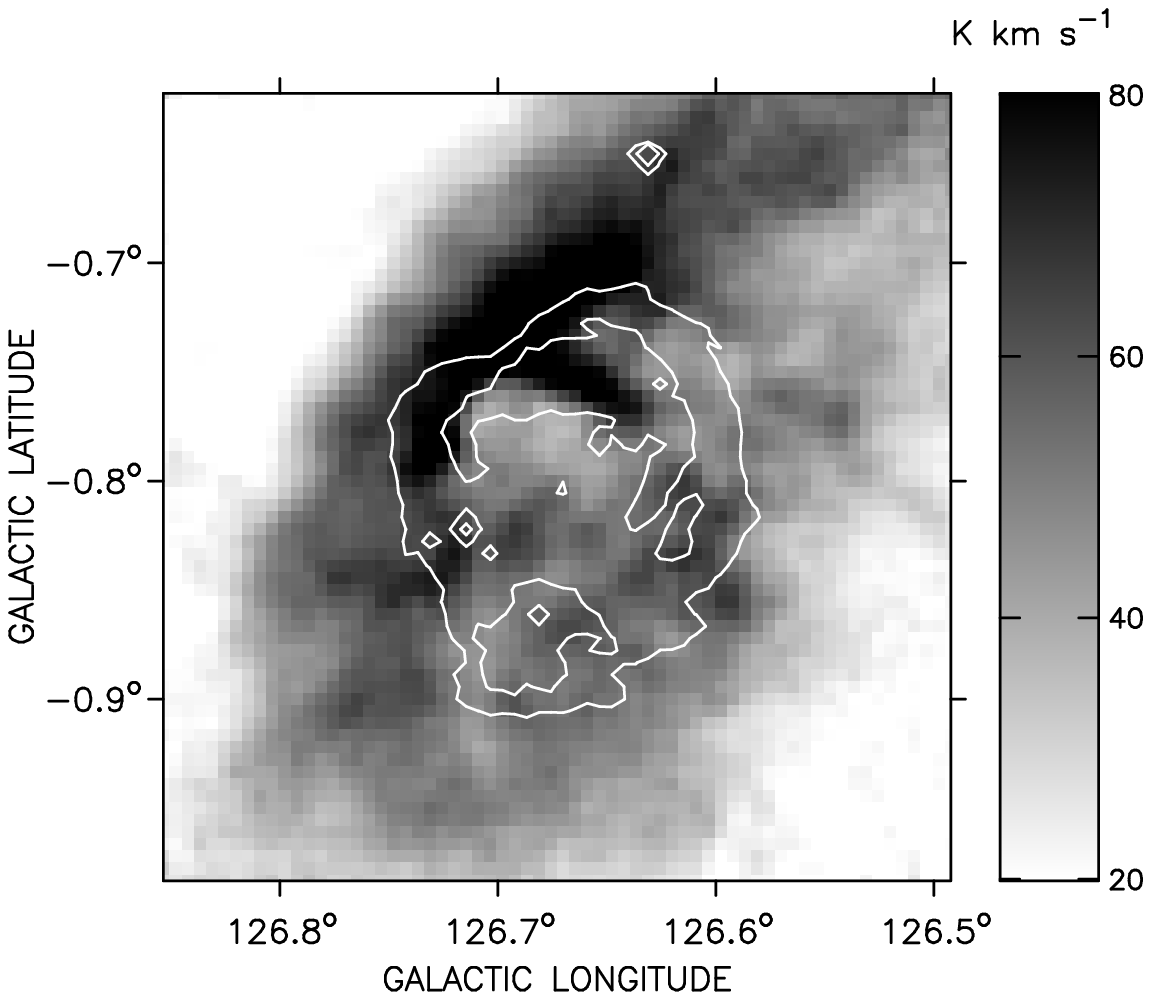}
\plotone{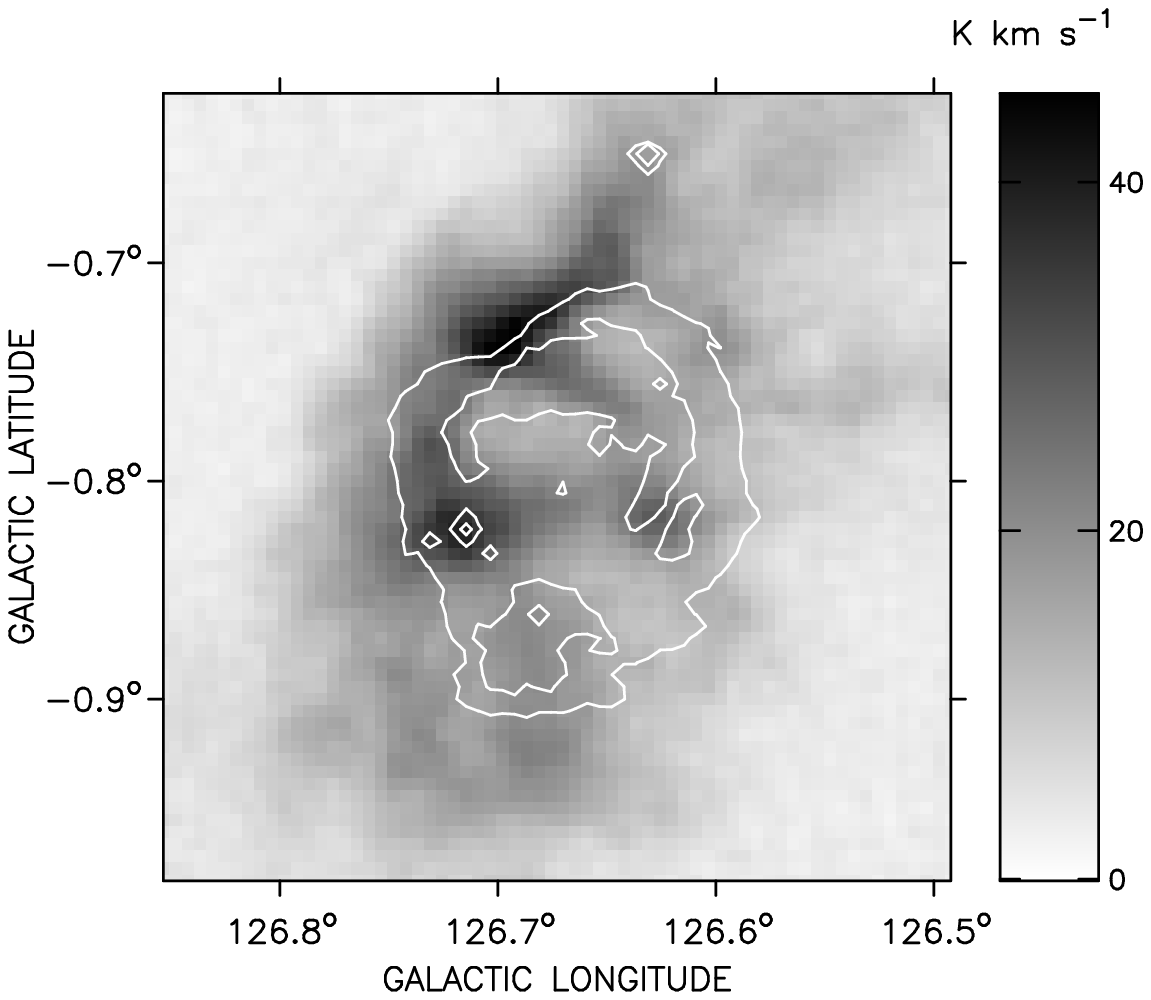}
%\plottwo{f17a.eps}{f17b.eps}
\caption{Top: integrated $^{12}$CO emission around KR~120. Bottom: integrated $^{13}$CO emission around KR~120. The white contours correspond to \emph{MSX} brightness levels of $3 \times 10^{-6}$, $1 \times 10^{-5}$, and $5 \times 10^{-5}$ W~m$^{-2}$~sr$^{-1}$ from Figure~\ref{figure-kr120-msxwithradiocontours}.}
\label{figure-kr120-12and13cowithmsxcontours}
\end{figure}

\begin{figure}
\plotone{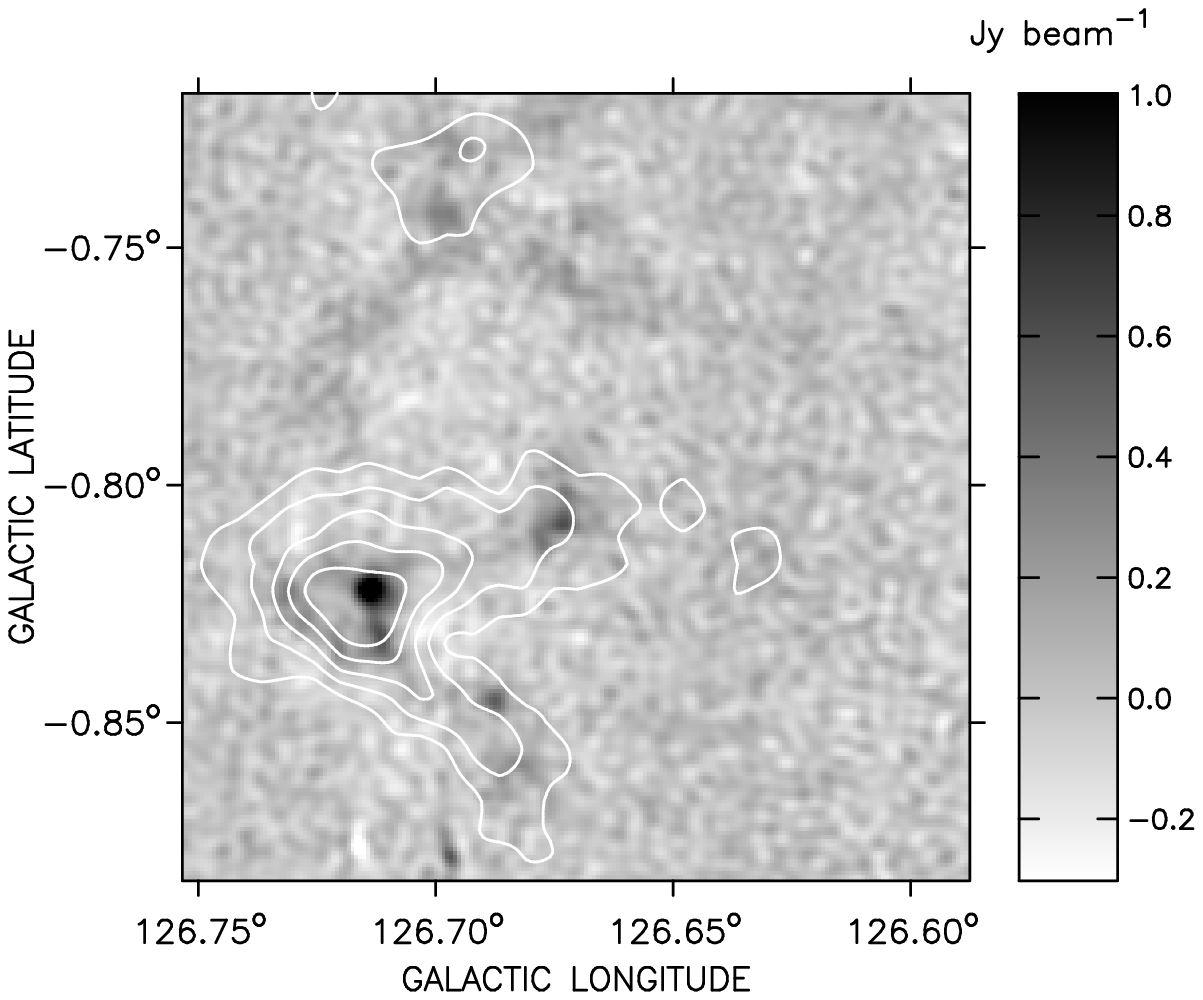}
\caption{Close up view of KR~120 in 850 $\mu$m. The white contours correspond to integrated C$^{18}$O brightness levels of 3, 4, 5, 6, and 7 K~km~s$^{-1}$.}
\label{figure-kr120-submmwithc18ocontours}
\end{figure}

For KR~120, the final 850 $\mu$m image is shown in
Figure~\ref{figure-kr120-finalsubmmimage}, and the CO emission associated with the \ion{H}{2} region has its peak around $-15$ km~s$^{-1}$ and is mainly concentrated on the Galactic east side (larger Galactic longitude), stretching on the outside of the region demarcated by the \emph{MSX} emission and further up toward the Galactic north above the \ion{H}{2} region. There is also a smaller concentration of CO emission to the Galactic southwest (smaller Galactic longitude). Figure~\ref{figure-kr120-12and13cowithmsxcontours} shows integrated CO emission from $^{12}$CO (left panel, integrated between $-18.0 < V_{\mathrm{LSR}} < -8.0$ km~s$^{-1}$) and $^{13}$CO (right panel, integrated between $-18.0 < V_{\mathrm{LSR}} < -10.0$ km~s$^{-1}$) overlaid with \emph{MSX} A-band contours corresponding to the feature seen in Figure~\ref{figure-kr120-msxwithradiocontours}. Evidently, even though $^{12}$CO and $^{13}$CO follow each other fairly well on large scales, the correlation appears to break down for the dense region close to the prominent KR~120 sub-mm source (discussed in the next paragraph), where $^{13}$CO and C$^{18}$O both peak but the $^{12}$CO does not (see Figure~\ref{figure-kr120-submmwithc18ocontours}). The $I_{\mathrm{peak}}$ of the integrated $^{13}$CO emission at the location of the sub-mm source has a value of $39.6 \pm 6.0$ K~km~s$^{-1}$, corresponding to a $^{13}$CO-based $N(\mathrm{H_2}) = (2.9 \pm 0.4) \times 10^{22}$ cm$^{-2}$ or $N(M) = 0.13 \pm 0.02$ g~cm$^{-2}$. Similarly, the C$^{18}$O-based column density is $N(\mathrm{H_2}) = (6.1 \pm 0.9) \times 10^{22}$ cm$^{-2}$ or $N(M) = 0.28 \pm 0.04$ g~cm$^{-2}$, fairly close to the estimate from $^{13}$CO. By summing over the entire $^{12}$CO feature seen in Figure~\ref{figure-kr120-12and13cowithmsxcontours}, the mass of the molecular material surrounding KR~120 is estimated to be $\sim 7600$ $M_{\sun}$.

The prominent KR~120 sub-mm source (JCMTSF J012332.0+614849) has a flux density of $1.6 \pm 0.4$ Jy within an aperture of $18 \arcsec$ around the peak. The deconvolved radius is $\sim 16 \arcsec$ and corresponds to a physical diameter of 0.2 pc at a distance of $1.44 \pm 0.26$ kpc. $B_{850}$ is 1.5 Jy~beam$^{-1}$ (S/N of $\sim23$). This corresponds to a dust-derived peak column density of $N(\mathrm{H_2}) = (7.0 \pm 1.4) \times 10^{22}$ cm$^{-2}$ or $N(M) = 0.33 \pm 0.07$ g~cm$^{-2}$. This dust-based column density is greater than the CO-based column density for the same location, again a possible opacity effect. The mass of the sub-mm source is $21 \pm 9$ $M_{\sun}$. For $T_d = 10$ K, the column density would increase to $2.0 \times 10^{23}$ cm$^{-2}$ or $N(M) = 0.95$ g~cm$^{-2}$ and the mass would increase to 60 $M_{\sun}$.

The SCUBA Legacy image for KR~120 looks very similar to Figure~\ref{figure-kr120-finalsubmmimage}. Apart from the main source, the SCUBA Legacy catalog contains 11 other sources distributed along the fainter 850 $\mu$m rims seen in Figure~\ref{figure-kr120-finalsubmmimage}. The main source is the most prominent one in the catalog with $\mathrm{S/N}= 40$, $F_{850} = 12.96$ Jy, and $B_{850} = 2.86 \pm 0.07$ Jy~beam$^{-1}$. The corresponding dust-derived peak column density is $N(\mathrm{H_2}) = (1.3 \pm 0.3) \times 10^{23}$ cm$^{-2}$ or $N(M) = 0.62 \pm 0.12$ g~cm$^{-2}$. The corresponding mass is $166 \pm 78$ $M_{\sun}$. This mass derived from the SCUBA Legacy catalog is $\sim 8$ times greater than the mass found in this work. This is mainly due to the larger area (effective radius of $58 \arcsec$) used in the catalog to obtain the flux, but also the higher peak $B_{850}$, and it illustrates the systematic uncertainty inherent in the method. We think all the remaining 11 sub-mm sources listed in the Legacy catalog are real detections as they have emission counterparts in our sub-mm maps (although at lower S/N) and are located within, or very close to, the C$^{18}$O contours shown in Figure~\ref{figure-kr120-submmwithc18ocontours}.

\begin{figure}
\plotone{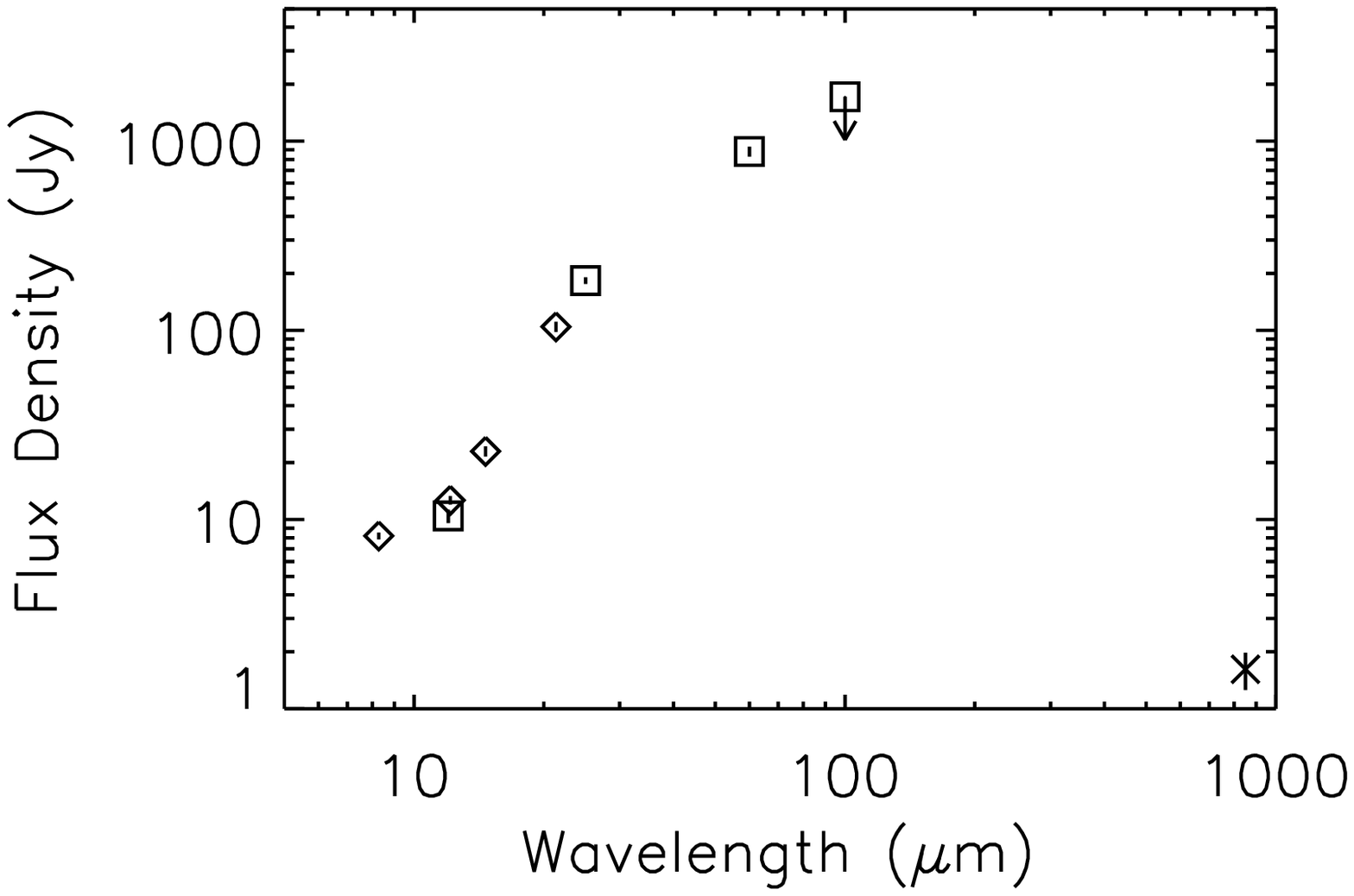}
\caption{Spectral energy distribution for the KR~120 850 $\mu$m source. \emph{IRAS} PSC values are squares, \emph{MSX6C} PSC values are diamonds and the 850 $\mu$m value is the cross. The vertical lines drawn on each symbol indicate $\pm 1\sigma$ uncertainty for each value. The arrow indicates an upper limit.}
\label{figure-kr120-sed}
\end{figure}

The main KR~120 850 $\mu$m source is spatially associated with IRAS 01202+6133 and MSX6C G126.7144-00.8220. In the \emph{IRAS} PSC, 12, 25, and 60 $\mu$m flux densities are listed as being good quality flux densities, while the 100 $\mu$m flux density only is an upper limit. The source is detected in all \emph{MSX} bands. Figure~\ref{figure-kr120-sed} shows the SED of this source from 8.3 to 850 $\mu$m. The arrow indicates an upper limit. Integrating under the curve yields an upper limit of $L_{IR} = 5600$ $L_{\sun}$, which is an order of magnitude higher than the other luminosities found in this study. The measured and calculated quantities are given in Table~\ref{table-results}.

\begin{figure}
\plotone{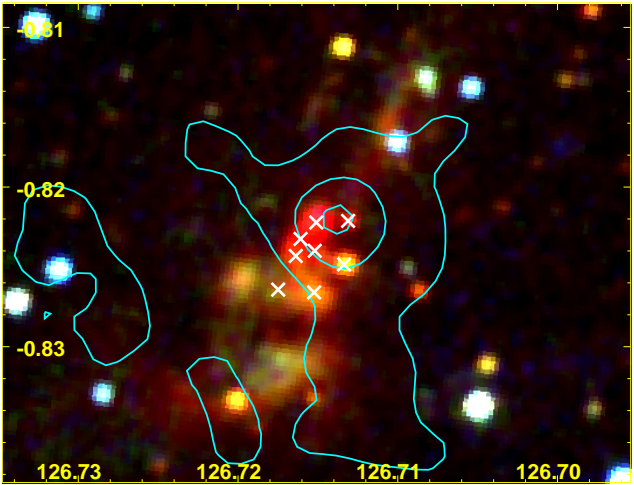}
\caption{$\sim 1\farcm8 \times 2\farcm3$ RGB composite of 2MASS $J$-band (Blue), $H$-band (Green) and $K_{\mathrm{s}}$-band (Red) images of KR~120. The contours correspond to 850 $\mu$m brightness levels of 0.2, 0.8, and 1.4 Jy~beam$^{-1}$.}
\label{figure-kr120-2massrgb}
\end{figure}

\begin{figure}
\epsscale{1.1}
\plotone{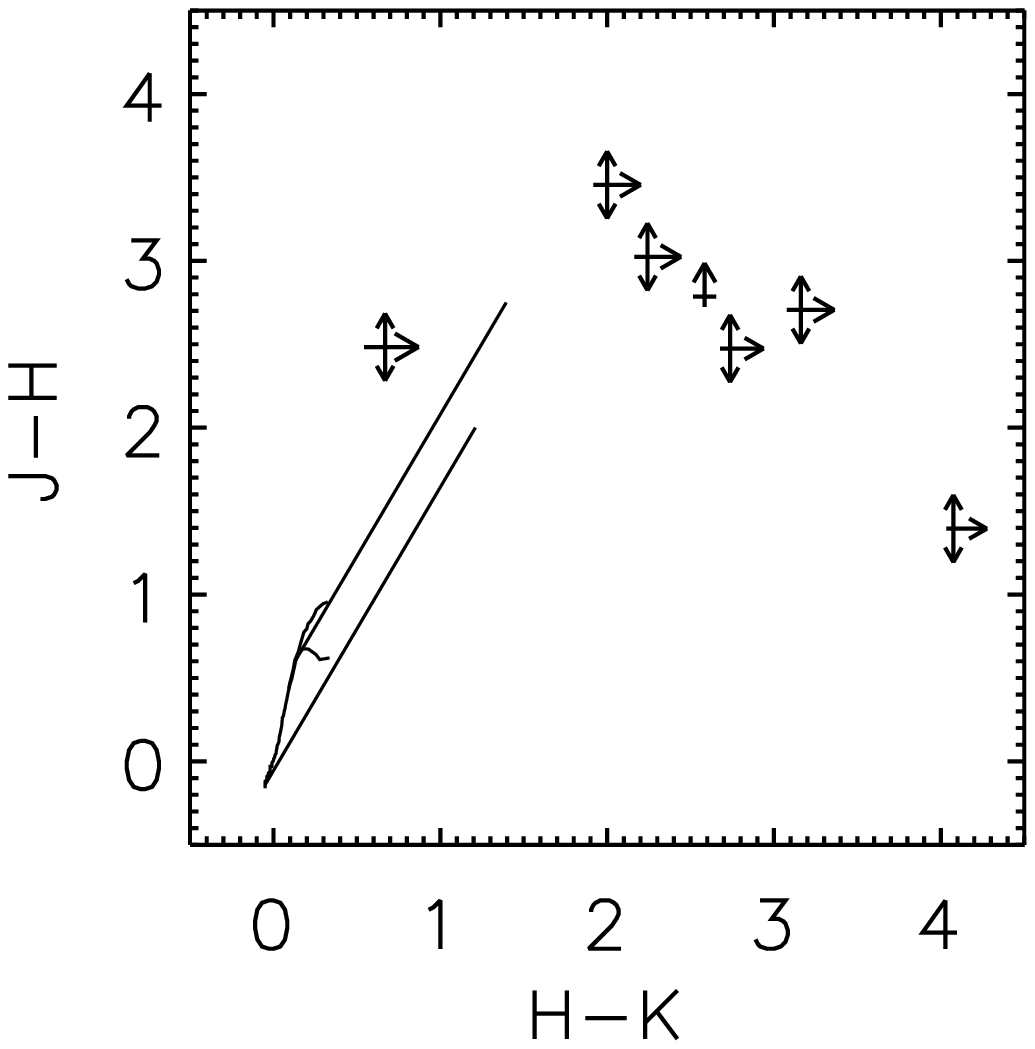}
\plotone{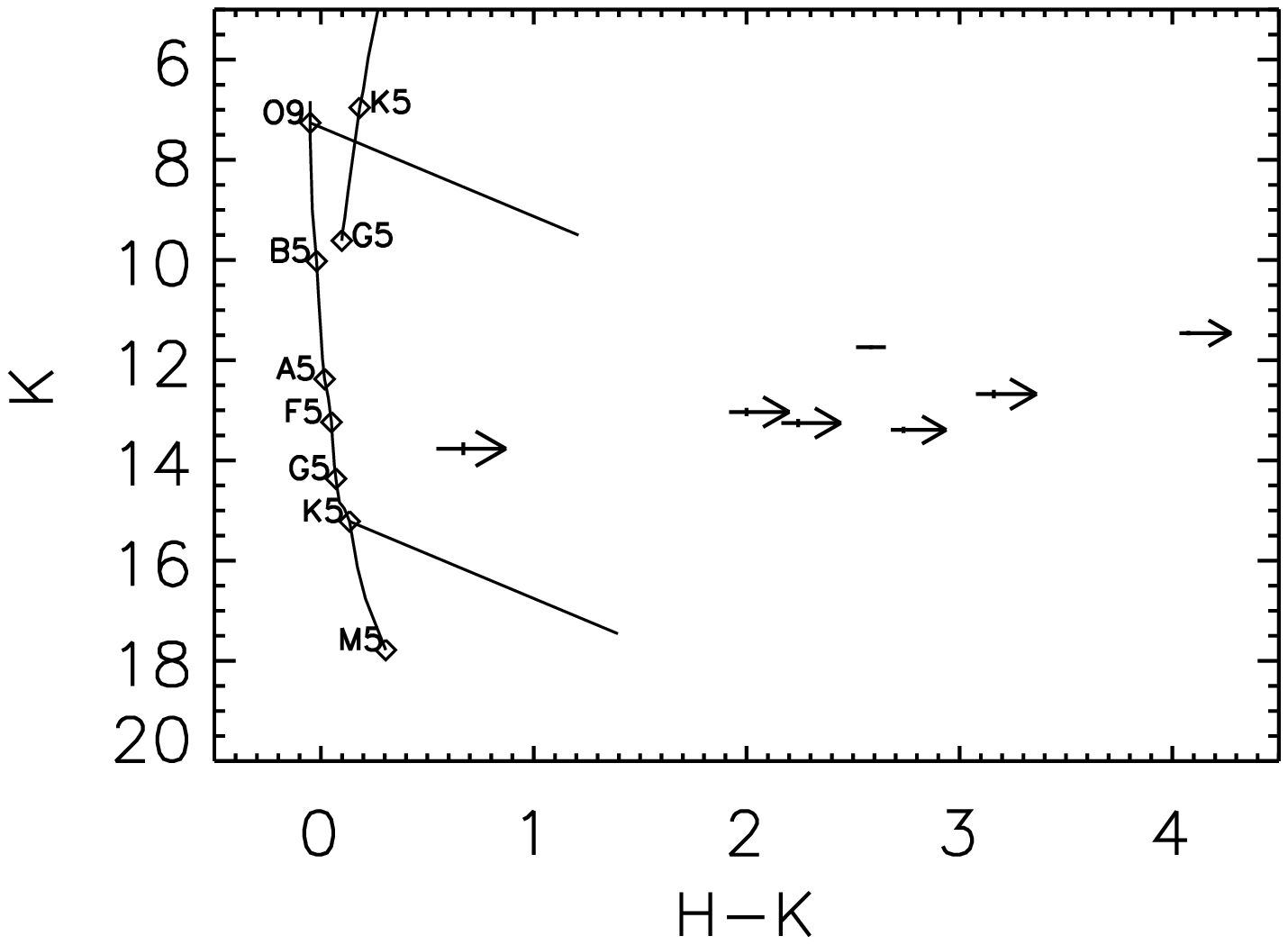}
\caption{Top: color-color diagram for the 2MASS sources within a radius of $0\farcm25$ of the KR~120 850 $\mu$m source. The solid lines in the lower left corner show the intrinsic colors of main-sequence (V) and giant (III) stars. The parallel lines show $A_V = 20$ reddening vectors for a K5~V and a O9~V star. Bottom: color-magnitude diagram for the same 2MASS sources at a distance of 1.44 kpc. The main-sequence and giant branch are shown with some representative spectral types. The parallel lines are again $A_V = 20$ reddening vectors. The size of the plotted crosses for the 2MASS sources indicates the uncertainty in 2MASS photometry. An arrow indicates sources where the 2MASS catalog only lists an upper brightness limit in the relevant band.}
\label{figure-kr120-2mass-colorcolor}
\epsscale{1.0}
\end{figure}

Figure~\ref{figure-kr120-2massrgb} shows a composite image of the region ($\sim 1\farcm8 \times 2\farcm3$) surrounding the dominant KR~120 850 $\mu$m source, made from 2MASS $J$-band (Blue), $H$-band (Green) and $K_{\mathrm{s}}$-band (Red) images. The contours, 850 $\mu$m brightness levels of 0.2, 0.8, and 1.4 Jy~beam$^{-1}$, trace the very central part of the sub-mm source, seen as the darkest small ``blob'' in Figures~\ref{figure-kr120-finalsubmmimage} and \ref{figure-kr120-submmwithc18ocontours}. A cluster of 2MASS sources are seen within the 850 $\mu$m contours, along with fuzzy NIR emission. The 2MASS catalog lists eight sources, all of which only have upper limit detections in the $J$-band, within a radius of $\sim 0\farcm25$ from the location of $B_{850}$, an area roughly corresponding to the extent of the 850 $\mu$m 0.2 Jy~beam$^{-1}$ contour. Figure~\ref{figure-kr120-2mass-colorcolor} shows these sources plotted in a 2MASS color-color diagram (top) and a color-magnitude diagram for a distance of 1.44 kpc (bottom), like Figure~\ref{figure-kr7-2mass-colorcolor}. All of the 2MASS sources in the color-color and color-magnitude diagrams have uncertain locations due to high uncertainty in 2MASS magnitudes, but at least almost all of them have to have $H - K \geq 2$, so the sources populate a region where highly embedded intermediate-mass YSOs would be. These 2MASS sources are likely to be a cluster of YSOs.

KR~120 has been the subject of previous studies. \citet{joncas-kr120multiwavelength-1992ApJ...387..591J} conducted a multiwavelength study of KR~120 in the optical, IR and radio, and found that KR~120 is an \ion{H}{2} region of age $\sim 2 \times 10^5$ years that is still enshrouded in the parental cloud. \citet{zavagno-kr120-1994A&A...281..491Z} discovered through spectroscopy a pre-main-sequence object (S~187H$\alpha$), located $\sim 2\farcm2$ from IRAS 01202+6133, and not covered by Figure~\ref{figure-kr120-2massrgb}. S~187H$\alpha$ is spatially located within JCMTSF J012313.2+614959 ($N(M) = 0.18 \pm 0.04$ g~cm$^{-2}$ and mass of $25 \pm 12$ $M_{\sun}$ from Table~\ref{table-scubalegacy}). \citet{salas-kr120-1998ApJ...500..853S} imaged the region near IRAS 01202+6133 (S~187 IR) in the NIR and discovered a curved molecular hydrogen outflow nearby that extends over a region of $76 \arcsec$ (0.55 pc at 1.44 kpc) that is probably due to a jet from a solar mass YSO; NIRS~1. NIRS~1 is spatially located within JCMTSF J012317.7+614735 ($N(M) = 0.14 \pm 0.03$ g~cm$^{-2}$ and mass of $7 \pm 3$ $M_{\sun}$ from Table~\ref{table-scubalegacy}) and is not covered by Figure~\ref{figure-kr120-2massrgb}. Furthermore, the \emph{MSX} source is listed as being associated with an OH maser, a bright signpost for a recently formed massive star \citep{argon-ohmasersforkr120-2000ApJS..129..159A}. The presence of the OH maser and pre-main-sequence objects in the neighborhood indicates that the investigated sub-mm clump is likely the site for ongoing star formation, something that agrees with the possible presence of an embedded cluster hinted at in Figures~\ref{figure-kr120-2massrgb} and \ref{figure-kr120-2mass-colorcolor}.

\section{Discussion}\label{sec-submmdiscussion}
The final list of sub-mm sources associated with the ring-like \ion{H}{2} regions includes the three main sources visually identified in our data along with an additional 10 sources for KR~7, 3 for KR~81, and 11 for KR~120 from the SCUBA Legacy catalog (see Table~\ref{table-scubalegacy} with the three main sources in bold). For consistency we will use the SCUBA Legacy catalog values of derived quantities in this discussion. The masses found for the regions fall within the range of sub-mm masses, $0.5-130$ $M_{\sun}$, that were found around the previously studied ring-like \ion{H}{2} region KR~140 \citep{kr140-submm-2001ApJ...552..601K}. The exception is the main source of KR~120 which is the most massive sub-mm source in the sample with 166 $M_{\sun}$. The visually determined sizes for the three prominent sources, 0.6, 0.4, and 0.2 pc, also fall within the range $0.2-0.7$ pc found for sources near KR~140. Around KR~140 as many as 22 sub-mm sources were found, while the three regions in this study have half or that or less. Given the noise level in the sub-mm images and the distances, a $3 \sigma$ source detection would correspond to a minimum mass sensitivity of $2-4$ $M_{\sun}$. Only 3 out of 22 sub-mm objects ($14\%$) for KR~140 have a mass $< 2$ $M_{\sun}$, so a large population of sub-mm sources is likely not missed in KR~7, KR~81, and KR~120. 

\begin{figure}
\epsscale{1.6}
\plotone{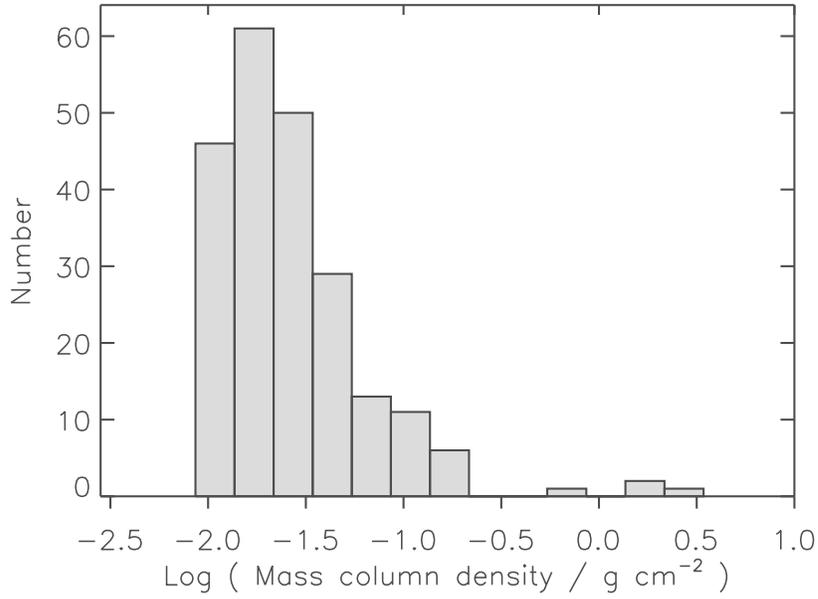}
\plotone{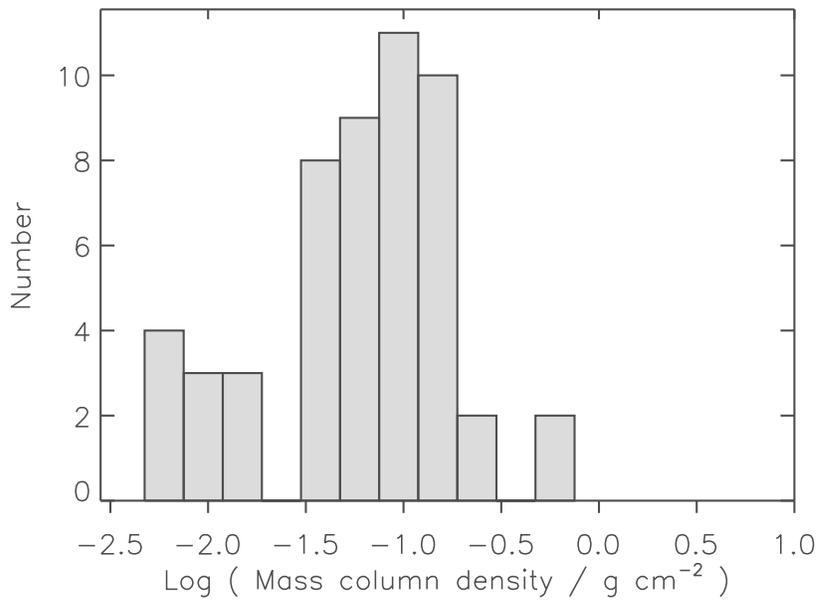}
\caption{Top: histogram of the 220 mass column densities for the HDL \citep{moore2007MNRAS.379..663M}. Bottom: histogram of the SCUBA Legacy mass column densities for KR~7, KR~81, and KR~120 from Table~\ref{table-scubalegacy} and for KR~140 from \citet{kr140-submm-2001ApJ...552..601K}. The bin size is 0.2 dex.}
\label{figure-hdlhistogram}
\end{figure}

\citet{moore2007MNRAS.379..663M} discovered 316 clumps in their 850 $\mu$m SCUBA study of the W3 GMC. For this study, we identify a subset of 220 clumps located in the high-density layer (HDL) of the W3 GMC. This layer runs parallel to the W4 \ion{H}{2} region and contains a number of luminous, massive star-forming regions (W3 Main, W3 (OH) and AFGL 333) that are likely to be examples of triggered star formation. Using the same dust temperature assumption as above in Section~\ref{sec-submmanalysis} ($T_d =20$ K) and a distance to W3 of 2.0 kpc, the peak mass column densities and sub-mm clump masses are computed. The distribution of peak mass column densities for the 220 HDL sub-mm sources is displayed in Figure~\ref{figure-hdlhistogram} (top). Peak mass column densities range between 0.009 g~cm$^{-2}$ and 3.1 g~cm$^{-2}$, with a median of 0.02 g~cm$^{-2}$ and an average of 0.07 g~cm$^{-2}$. The distribution of peak mass column densities for KR~7, KR~81 and KR~120 (from Table~\ref{table-scubalegacy}), and for KR~140 \citep{kr140-submm-2001ApJ...552..601K} is shown in Figure~\ref{figure-hdlhistogram} (bottom). The values found for the three prominent KR objects in this study, $0.16 - 0.62$ g~cm$^{-2}$, land above both the median and average value associated with the HDL. Interestingly, while there is overlap in the two mass column density distributions, the HDL distribution is very heavily weighted to lower mass column densities compared to the ring-like \ion{H}{2} region distribution. It should be noted though that the maximum peak mass column density for the HDL sources is about an order of magnitude larger than found surrounding the ring-like \ion{H}{2} regions.

\begin{figure}
\plotone{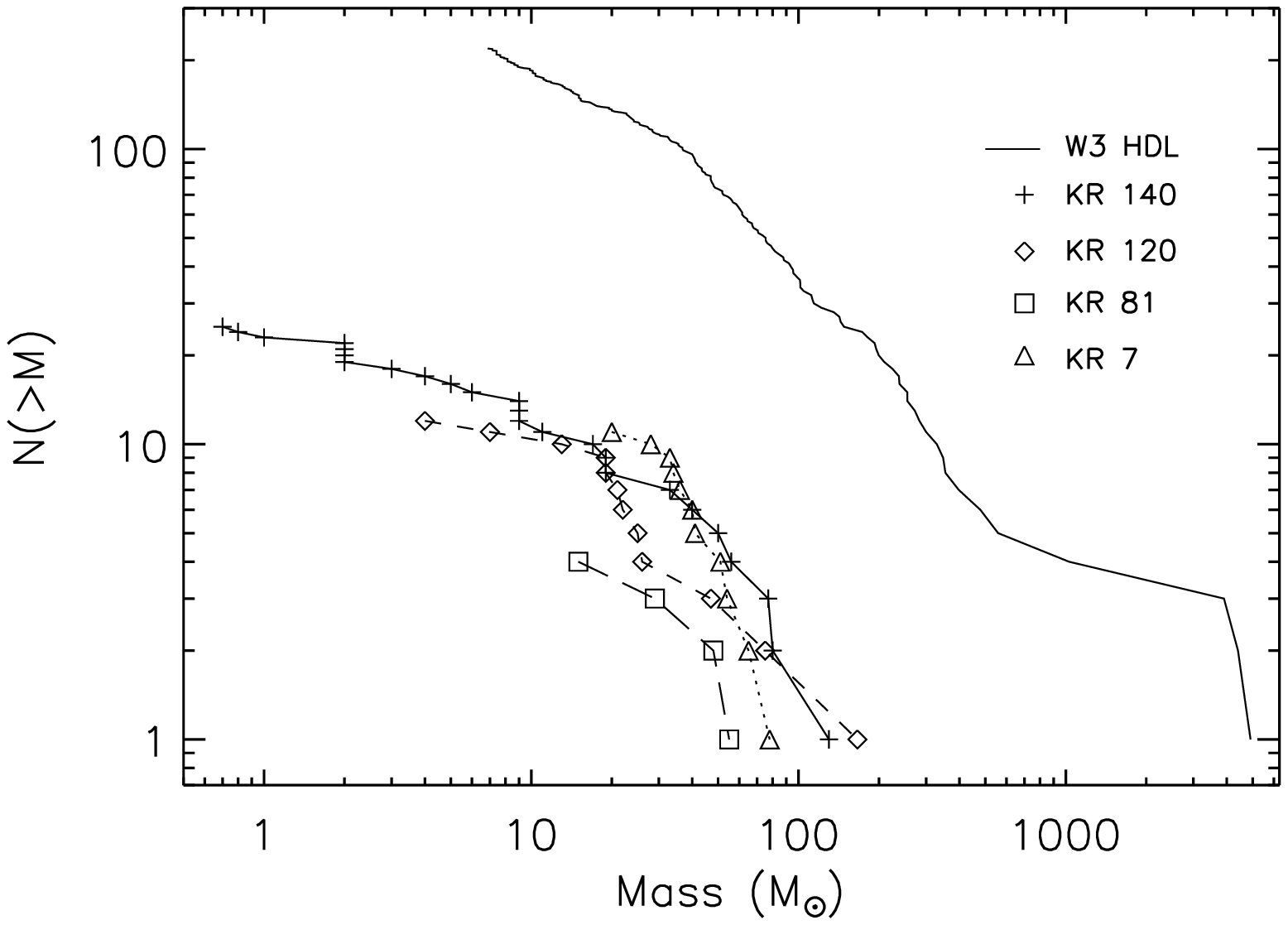}
\caption{Cumulative plot of sub-mm source mass for the HDL \citep[solid line, from][]{moore2007MNRAS.379..663M}, KR~7 (triangles), KR~81 (squares), KR~120 (diamonds), and KR~140 \citep[crosses, from][]{kr140-submm-2001ApJ...552..601K}.}
\label{figure-cumulmass}
\end{figure}

The masses of the 220 sub-mm sources in the HDL range between 7 and 4900 $M_{\sun}$, with a median of 31 and an average of 123 $M_{\sun}$. The sub-mm masses found in this study fall within the range of masses and around the median of masses associated with the HDL, but the maximum masses found within the HDL are at least an order of magnitude higher than found in this study. Figure~\ref{figure-cumulmass} shows a cumulative plot of sub-mm source mass for the HDL (solid line), KR~7 (triangles), KR~81 (squares), KR~120 (diamonds), and KR~140 \citep[crosses; from][]{kr140-submm-2001ApJ...552..601K}. Table~\ref{table-massfits} shows the indexes ($\alpha$) from fitting power laws of the form $N(>M) = N_0 M^{-\alpha}$ to the mass distributions, as well as the fraction of the total mass contained within the first $50\%$ of objects, counting from the smallest to the highest masses. For comparison, a Salpeter (stellar-like) mass distribution would have an index of $\alpha = 1.35$ and the first $50\%$ of objects would have about $18\%$ of the total mass. Except for KR~7, all of the regions we examine have clump mass
distributions that are consistent with the typical clump structure
index of $0.6 \pm 0.1$ associated with molecular cloud structure \citep{kramer1998AA...329..249K,williams2000prpl.conf...97W}. The KR~7 clumps are an interesting
outlier as they have a very steep index (the first $50 \%$ of the sources contain about $36 \%$ of the total mass) more typical of a stellar
initial mass function (IMF) that is usually only seen in smaller (1000 AU scale) structure within
molecular clouds \citep[e.g.,][]{motte1998AA...336..150M}. However, the KR~7 clumps also span the most limited range of masses of any of the investigated regions.

The respective masses for the surrounding molecular material are $3600$ $M_{\sun}$ within a rough diameter of 19 pc (KR~7), $1900$ $M_{\sun}$ within 12 pc (KR~81), and $7600$ $M_{\sun}$ within 13 pc (KR~120). \citet{kr140-2nd-2000ApJ...539..283B} found that the molecular cloud surrounding KR~140 contained about 5000 $M_{\sun}$. Data from the $^{12}$CO FCRAO Outer Galaxy Survey \citep{ogs-1998ApJS..115..241H} with a spectral resolution of $0.98$ km~s$^{-1}$ and an angular resolution of $46 \arcsec$ show that the molecular material seen around KR~120 and KR~140 is likely part of larger molecular clouds (subtending $1\degr-2 \degr$ on the sky). The material around KR~7 and KR~81 does not show connections to larger molecular clouds. This makes at least two examples of massive stars forming in clouds with masses $< 10^4$ $M_{\sun}$, which is quite unexpected. Massive stars are expected to form in giant molecular clouds (GMCs), which range $10^4 - 10^6$ $M_{\sun}$ \citep{ladalada-2003ARAA..41...57L}.

\begin{figure}
\plotone{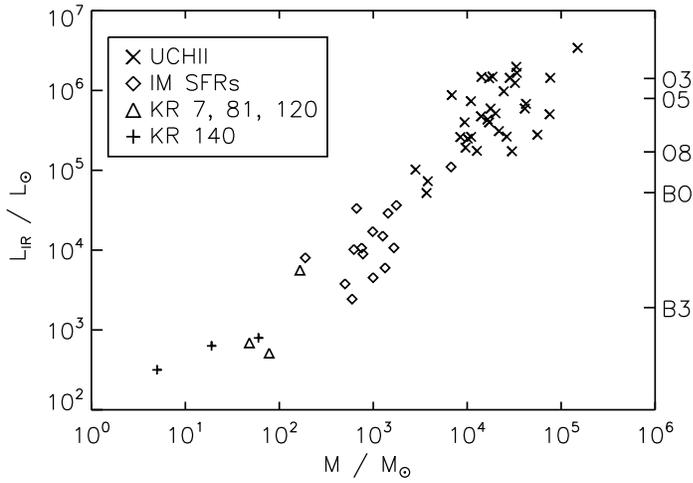}
\caption{Luminosity vs. associated mass. IM SFRs (15 objects) are diamonds, and UC\ion{H}{2} regions (33 objects) are crosses \citep[from][]{arvidsson-imsfrs}. The three prominent KR sub-mm sources are triangles. KR~140 sources are taken from \citet{kr140-submm-2001ApJ...552..601K}. The scale on the right is spectral type of a single class V star.}
\label{figure-luminosityvsmass}
\end{figure}

Figure~\ref{figure-luminosityvsmass} shows total integrated luminosity, $L_{IR}$, as a function of associated clump mass for three types of objects. Ultra-compact \ion{H}{2} (UC\ion{H}{2}) regions, representing sites of massive star formation are crosses, while intermediate-mass star-forming regions (IM SFRs) are diamonds, both taken from \citet{arvidsson-imsfrs}. The three prominent KR sub-mm sources are triangles. Three sources around KR~140 where the identification of infrared sources (IRAS 02174+6052, IRAS 02171+6058 and IRAS 021757+6053) corresponding to sub-mm sources (KMJB 1, KMJB 3 and KMJB 17, 18 and 19) is relatively straightforward is taken from \citet{kr140-submm-2001ApJ...552..601K}. It still shows a correlation ($r^2 = 0.89$) between luminosity and associated clump mass for IM SFRs, UC\ion{H}{2} regions and KR sub-mm sources. The data are fit by $L \propto M^{1.04 \pm 0.05}$ which largely agrees with the similar value found for UC\ion{H}{2} regions and IM SFRs in \citet{arvidsson-imsfrs}, and with the value found for compact \ion{H}{2} regions in \citet{chini1987A&A...181..378C}. From Figure~\ref{figure-luminosityvsmass} it appears the relationship holds over four orders of magnitude in mass, extending another order of magnitude to lower mass less than found in \citet{arvidsson-imsfrs} and two orders of magnitude lower in mass than shown in \citet{chini1987A&A...181..378C}. This is significant because a constant ratio of luminosity to associated mass is what is expected if star formation follows the same IMF in all star forming regions and that the star formation efficiency for converting gas into stars is the same over many orders of magnitude in associated mass \citep{chini1987A&A...181..378C}.

\section{Conclusions}\label{sec-submmconclusions}
Three ring-like morphology \ion{H}{2} regions in the outer Galaxy, KR~7, KR~81, and KR~120 (Sh 2-124, Sh 2-165, and Sh 2-187), have been investigated using SCUBA 850 $\mu$m observations and molecular line observations ($^{12}$CO, $^{13}$CO and C$^{18}$O). They are found to each have one dominant 850 $\mu$m source, located in the interface region between the \ion{H}{2} region and the surrounding molecular material. At the same location as these dominant 850 $\mu$m sources, peaks are found in the integrated molecular spectral maps, confirming these as locations of cold, dense material. Estimating the peak mass column densities toward the dominant sources results in values of $0.1 - 0.6$ g~cm$^{-2}$, comparable to the peak mass column densities found for IM SFRs \citep{arvidsson-imsfrs}. The clump masses associated with the three dominant sources, estimated to be 51, 27, and 21 $M_{\sun}$, fall within the range of clump masses previously found for KR~140 by \citet{kr140-submm-2001ApJ...552..601K}. The same is true for the sizes, 0.6, 0.4, and 0.2 pc, that fall within the range $0.2-0.7$ pc. Using 2MASS photometry, a possible embedded cluster of YSOs are found within the dominant sub-mm source of KR~120. This is likely a site for ongoing star formation, consistent with earlier studies that found signatures of star formation in the immediate vicinity. Candidates for embedded YSOs within the sub-mm sources are found for KR~7, while sources toward KR~81 can be explained as either foreground or background objects. Using the SCUBA Legacy catalog, the three \ion{H}{2} regions are found to be less populated with sub-mm sources than the previously studied \ion{H}{2} region KR~140. All but one of the regions investigated have sub-mm source mass distributions described by shallow sloped (sub-Salpeter) power laws. The exception, KR~7, has a very steep power law fit but covers a much more limited mass range.

\begin{acknowledgements}
The authors thank C. Brunt for providing the CO data. This publication makes use of the NASA/IPAC Infrared Science Archive, which is
operated by the Jet Propulsion Laboratory, California
Institute of Technology, under contract with the National
Aeronautics and Space Administration. This publication makes use of data products from the Two Micron All Sky Survey, which is a joint project of the University of Massachusetts and the Infrared Processing and Analysis Center/California Institute of Technology, funded by the National Aeronautics and Space Administration and the National Science Foundation.
\end{acknowledgements}

\clearpage

\begin{deluxetable}{lllrrcc}
\tablecolumns{7}
\tabletypesize{\footnotesize}
\tablewidth{0pc}
\tablecaption{Table of the Investigated Regions.\label{table-KRsubmmobservations}}
\tablehead{
\colhead{Region} & \colhead{R.A.} & \colhead{Decl.} & \colhead{$\ell$} & \colhead{$b$} & \colhead{Distance}& \colhead{Designations}\\
\colhead{} & \colhead{(h m s)} & \colhead{($\degr$ $\arcmin$ $\arcsec$)}   & \colhead{($\degr$)}    & \colhead{($\degr$)} & \colhead{(kpc)} & \colhead{}
}
\startdata
KR~7 & 21 38 17.0 & +50 19 48 & 94.461 & $-1.549$ & $2.8 \pm 0.4^a$ & Sh 2-124, LBN 426\\
KR~81 & 23 39 43.7 & +61 54 58 & 114.600 & 0.210 & $1.9 \pm 0.4^a$ & Sh 2-165, LBN 565\\
KR~120 & 01 22 58.0 & +61 48 16 & 126.647 & $-0.840$ & $1.44 \pm 0.26^b$ & Sh 2-187, LBN 630\\
\enddata
\tablecomments{Coordinates are J2000.\\
$^a$From \citet{kr7and81distance-2003ApJ...598.1005F}.\\
$^b$From \citet{kr120distance-2007A&A...470..161R}.}
\end{deluxetable}

\begin{deluxetable}{lccc}
\tablecolumns{4}
\tabletypesize{\footnotesize}
\tablewidth{0pc}
\tablecaption{Results of Analysis of the Three Main 850 $\mu$m Sources.\label{table-results}}
\tablehead{
\colhead{Region} & \colhead{KR~7} & \colhead{KR~81}   & \colhead{KR~120}\\
\colhead{JCMTSF} &  \colhead{J213822.5+501908} &  \colhead{J233917.8+615914} &  \colhead{J012332.0+614849}}
\startdata
$I_{\mathrm{peak}}^a$ (K~km~s$^{-1}$) & $7.9 \pm 1.2$ & $13.5 \pm 2.0$ & $39.6 \pm 6.0$\\
$I_{\mathrm{peak}}^b$ (K~km~s$^{-1}$) & \nodata & $1.3 \pm 0.2$ & $8.1 \pm 1.2$\\
Radius ($\arcsec$) & 24 & 24 & 16\\
$F_{850}$ (Jy) & $1.1 \pm 0.2$ & $1.2 \pm 0.3$ & $1.6 \pm 0.4$\\
$B_{850}$ (Jy~beam$^{-1}$) & $0.50 \pm 0.11$ & $0.48 \pm 0.12$ & $1.52 \pm 0.31$\\
$N(M)_{\mathrm{peak}}^a$ (g~cm$^{-2}$) & $0.026 \pm 0.004$ & $0.045 \pm 0.007$ & $0.13 \pm 0.02$\\
$N(M)_{\mathrm{peak}}^b$ (g~cm$^{-2}$) &  \nodata & $0.045 \pm 0.007$ & $0.28 \pm 0.04$\\
$N(M)_{\mathrm{peak}}^c$ (g~cm$^{-2}$) & $0.11 \pm 0.02$ & $0.10 \pm 0.03$ & $0.33 \pm 0.07$\\
Diameter (pc) & 0.6 & 0.4 & 0.2\\
Mass$^c$ ($M_{\sun}$) & $51 \pm 19$ & $27 \pm 13$ & $21 \pm 9$\\
\tableline
$N(M)_{\mathrm{peak}}^d$ (g~cm$^{-2}$) & $0.16 \pm 0.03$ & $0.18 \pm 0.04$ & $0.62 \pm 0.12$\\
Mass$^d$ ($M_{\sun}$) & $78 \pm 32$ & $48 \pm 25$ & $166 \pm 78$\\
\tableline
MSX6C & G094.4674-01.5705 & G114.5696+00.2899 & G126.7144-00.8220\\
$F_{\mathrm{A}}$ (Jy) & $0.554 \pm 4.4\%$ & $2.60 \pm 4.1\%$ & $8.21 \pm 4.1\%$\\
$F_{\mathrm{C}}$ (Jy) & \nodata & $3.83 \pm 5.2\%$ & $12.6 \pm 5.0\%$\\
$F_{\mathrm{E}}$ (Jy) & \nodata & $5.47 \pm 6.2\%$ & $105 \pm 6.0\%$\\
\tableline
\emph{IRAS} &  21366+5005 &  23369+6142 &  01202+6133\\
$F_{12}$ (Jy) & $0.94 \pm 5\%$ & $5.025 \pm 8\%$ & $10.44 \pm 8\%$\\
$F_{25}$ (Jy) & $1.60 \pm 7\%$ & $8.322 \pm 6\%$ & $182.3 \pm 4\%$\\
$F_{60}$ (Jy) & $14.0 \pm 18\%$ & $37.5 \pm 12\%$ & $881.5 \pm 6\%$\\
$F_{100}$ (Jy) & $\leq66.3$ & $153.8 \pm 18\%$ & $\leq1716$\\
$L_{\mathrm{IR}}$ ($L_{\sun}$) & $\leq500$ & $700 \pm 300$ & $\leq5600$\\
\enddata
\tablecomments{$^a$Based on $^{13}$CO.\\
$^b$Based on C$^{18}$O.\\
$^c$Based on 850 $\mu$m.\\
$^d$From Table~\ref{table-scubalegacy}.}
\end{deluxetable}

\clearpage

\begin{deluxetable}{lcrrcrrcc}
\tablecolumns{9}
\tabletypesize{\footnotesize}
\tablewidth{0pc}
\tablecaption{SCUBA Legacy 850 $\mu$m Sources.\label{table-scubalegacy}}
\tablehead{
\colhead{Region} & \colhead{Source} & \colhead{$\ell$} & \colhead{$b$} & \colhead{$B_{850}$} & \colhead{S/N}   & \colhead{$F_{850}$} & \colhead{$N(M)_{\mathrm{peak}}$}   &\colhead{Mass}\\
\colhead{} & \colhead{JCMTSF} & \colhead{($\degr$)}    & \colhead{($\degr$)} & \colhead{(Jy~beam$^{-1}$)} & \colhead{} & \colhead{(Jy)} & \colhead{(g~cm$^{-2}$)}    & \colhead{($M_{\sun}$)}
}
\startdata
\textbf{KR~7} & \textbf{J213822.5+501908} & $\mathbf{94.4642}$ & $\mathbf{-1.5675}$ & \textbf{0.73} & \textbf{13.0} & \textbf{1.62} & $\mathbf{0.16 \pm 0.03}$ & $\mathbf{78 \pm 32}$ \\
KR~7 & J213825.8+502714 & $94.5605$ & $-1.4724$ & 0.25 & 4.4 & 1.35 & $0.05 \pm 0.02$ & $65 \pm 27$ \\
KR~7 & J213801.8+501956 & $94.4319$ & $-1.5209$ & 0.23 & 4.2 & 0.58 & $0.05 \pm 0.01$ & $28 \pm 12$ \\
KR~7 & J213756.2+502344 & $94.4630$ & $-1.4637$ & 0.23 & 4.1 & 0.84 & $0.05 \pm 0.02$ & $41 \pm 17$ \\
KR~7 & J213749.9+502102 & $94.4205$ & $-1.4862$ & 0.23 & 4.1 & 0.83 & $0.05 \pm 0.02$ & $40 \pm 17$ \\
KR~7 & J213813.7+501908 & $94.4467$ & $-1.5519$ & 0.21 & 3.9 & 1.06 & $0.05 \pm 0.01$ & $51 \pm 21$ \\
KR~7 & J213753.1+502308 & $94.4502$ & $-1.4657$ & 0.21 & 3.7 & 0.75 & $0.05 \pm 0.02$ & $36 \pm 15$ \\
KR~7 & J213835.7+502137 & $94.5180$ & $-1.5599$ & 0.19 & 3.3 & 0.41 & $0.04 \pm 0.02$ & $20 \pm 8$ \\
KR~7 & J213803.0+501850 & $94.4221$ & $-1.5367$ & 0.19 & 3.5 & 0.71 & $0.04 \pm 0.01$ & $34 \pm 14$ \\
KR~7 & J213759.9+501908 & $94.4193$ & $-1.5275$ & 0.19 & 3.5 & 0.68 & $0.04 \pm 0.01$ & $33 \pm 14$ \\
KR~7 & J213749.3+502202 & $94.4304$ & $-1.4727$ & 0.19 & 3.3 & 1.12 & $0.04 \pm 0.02$ & $54 \pm 22$ \\
\textbf{KR~81} & \textbf{J233917.8+615914} & $\mathbf{114.5707}$ & $\mathbf{0.2923}$ & \textbf{0.84} & \textbf{13.8} & \textbf{2.17} & $\mathbf{0.18 \pm 0.04}$ & $\mathbf{48 \pm 25}$ \\
KR~81 & J233922.3+615702 & $114.5691$ & $0.2546$ & 0.46 & 7.6 & 2.47 & $0.10 \pm 0.02$ & $55 \pm 28$ \\
KR~81 & J233923.7+615938 & $114.5837$ & $0.2955$ & 0.36 & 5.9 & 0.67 & $0.08 \pm 0.02$ & $15 \pm 8$ \\
KR~81 & J233912.1+615649 & $114.5489$ & $0.2566$ & 0.25 & 4.1 & 1.29 & $0.05 \pm 0.02$ & $29 \pm 15$ \\
\textbf{KR~120} & \textbf{J012332.0+614849} & $\mathbf{126.7127}$ & $\mathbf{-0.8223}$ & \textbf{2.86} & \textbf{40.0} & \textbf{12.96} & $\mathbf{0.62 \pm 0.12}$ & $\mathbf{166 \pm 78}$ \\
KR~120 & J012338.8+614819 & $126.7270$ & $-0.8289$ & 1.16 & 16.2 & 5.83 & $0.25 \pm 0.05$ & $75 \pm 35$ \\
KR~120 & J012329.0+615336 & $126.6969$ & $-0.7439$ & 0.86 & 12.1 & 3.71 & $0.19 \pm 0.04$ & $47 \pm 22$ \\
KR~120 & J012313.2+614959 & $126.6736$ & $-0.8076$ & 0.82 & 11.5 & 1.93 & $0.18 \pm 0.04$ & $25 \pm 12$ \\
KR~120 & J012317.7+614735 & $126.6873$ & $-0.8462$ & 0.67 & 9.7 & 0.58 & $0.14 \pm 0.03$ & $7 \pm 3$ \\
KR~120 & J012328.9+615424 & $126.6951$ & $-0.7307$ & 0.61 & 8.3 & 1.75 & $0.13 \pm 0.03$ & $22 \pm 11$ \\
KR~120 & J012318.8+615312 & $126.6779$ & $-0.7530$ & 0.48 & 6.8 & 1.49 & $0.10 \pm 0.03$ & $19 \pm 9$ \\
KR~120 & J012338.5+615201 & $126.7187$ & $-0.7678$ & 0.44 & 6.2 & 1.45 & $0.09 \pm 0.02$ & $19 \pm 9$ \\
KR~120 & J012342.8+615114 & $126.7287$ & $-0.7797$ & 0.42 & 5.9 & 2.04 & $0.09 \pm 0.02$ & $26 \pm 12$ \\
KR~120 & J012250.4+614951 & $126.6293$ & $-0.8153$ & 0.38 & 5.5 & 1.61 & $0.08 \pm 0.02$ & $21 \pm 10$ \\
KR~120 & J012321.2+615454 & $126.6791$ & $-0.7243$ & 0.38 & 4.9 & 1.00 & $0.08 \pm 0.02$ & $13 \pm 6$ \\
KR~120 & J012313.6+614653 & $126.6807$ & $-0.8588$ & 0.36 & 5.0 & 0.34 & $0.08 \pm 0.02$ & $4 \pm 2$ \\
\enddata
\tablecomments{Final sample of sub-mm sources considered for comparisons. The three main sources are shown in bold. Columns $2-7$ are taken from \citet{scubalegacy-2008ApJS..175..277D}. Column 8 is calculated using an absolute flux uncertainty of $20 \%$. Column 9 is calculated using a flux density uncertainty of $30 \%$.}
\end{deluxetable}

\begin{deluxetable}{lccc}
\tablecolumns{4}
\tabletypesize{\footnotesize}
\tablewidth{0pc}
\tablecaption{Mass Distribution Characteristics.\label{table-massfits}}
\tablehead{
\colhead{Region} & \colhead{$\alpha$} & \colhead{$50\%$ Mass Values} & \colhead{$N$}
}
\startdata
KR~7 & $1.8 \pm 0.2$ & $36\%$ & 11\\
KR~81 & $0.9 \pm 0.3$ & $30\%$ & 4\\
KR~120 & $0.72 \pm 0.08$ & $19\%$ & 12\\
KR~140 & $0.50 \pm 0.04$ & $7\%$ & 22\\
W3~HDL & $0.76 \pm 0.01$ & $6\%$ & 220
\enddata
\end{deluxetable}

\clearpage

\clearpage

\clearpage

\clearpage

\clearpage
\epsscale{1.1}

\clearpage

\epsscale{2.0}

\clearpage

\clearpage
\epsscale{1.3}

\clearpage

\epsscale{1.1}

\clearpage

\epsscale{2.0}

\clearpage

\clearpage

\epsscale{1.3}

\clearpage
\epsscale{1.1}

\clearpage

\clearpage

\epsscale{1.3}

\end{document}